\documentclass[11pt,aps,showpacs,nofootinbib,prd,aps,epsf,floats,rotate]{revtex4}
\usepackage{amsmath, amssymb, graphics}

\newcommand{\mathsym}[1]{{}}

\usepackage{graphicx}
\usepackage{amsmath}
\usepackage{amssymb}
\usepackage{bm}
\newcommand{\be}{\begin{equation}}
\newcommand{\ee}{\end{equation}}
\newcommand{\bea}{\begin{eqnarray}}
\newcommand{\eea}{\end{eqnarray}}

\newcommand{\rem}[1]{}
\newsavebox{\PSLASH}
 \sbox{\PSLASH}{$p$\hspace{-1.8mm}/}
 
\renewcommand{\theequation}{\thesection.\arabic{equation}}
\newcounter{saveeqn}
\newcommand{\add}{\addtocounter{equation}{1}}
\newcommand{\alpheqn}{\setcounter{saveeqn}{\value{equation}}%
\setcounter{equation}{0}%
\renewcommand{\theequation}{\mbox{\thesection.\arabic{saveeqn}{\alph{equation}}}}}
\newcommand{\reseteqn}{\setcounter{equation}{\value{saveeqn}}%
\renewcommand{\theequation}{\thesection.\arabic{equation}}}
\newenvironment{nedalph}{\add\alpheqn\begin{eqnarray}}{\end{eqnarray}\reseteqn}
 \newsavebox{\notrightarrow}
 \sbox{\notrightarrow}{$\to$\hspace{-4mm}/}
 
 \newsavebox{\PARTIALSLASH}
 \sbox{\PARTIALSLASH}{$\partial$\hspace{-1.6mm}/}
 
 \newsavebox{\ASLASH}
 \sbox{\ASLASH}{$A$\hspace{-2.1mm}/}
 
 \newsavebox{\KSLASH}
 \sbox{\KSLASH}{$k$\hspace{-1.8mm}/}
 
 \newsavebox{\LSLASH}
 \sbox{\LSLASH}{$\ell$\hspace{-1.8mm}/}
 
 \newsavebox{\QSLASH}
 \sbox{\QSLASH}{$q$\hspace{-1.8mm}/}
 
 \newsavebox{\DSLASH}
 \sbox{\DSLASH}{$D$\hspace{-2.2mm}/}
 
 \newsavebox{\DbfSLASH}
 \sbox{\DbfSLASH}{${\mathbf D}$\hspace{-2.8mm}/}
 
 \newsavebox{\DELVECRIGHT}
 \sbox{\DELVECRIGHT}{$\stackrel{\rightarrow}{\partial}$}
 
 \newcommand{\blue}{\IfColor{\textCadetBlue}{}}
\newcommand{\black}{\IfColor{\textBlack}{}}
\newcommand{\red}{\IfColor{\textRed}{}}
\newcommand{\green}{\IfColor{\textOliveGreen}{}}
\newcommand{\lila}{\IfColor{\textRedViolet}{}}

\begin{document}

\begin{flushright}
SUT-P-2007/001b\\
IPM/P-2007/037\\ arXiv:0705.4384 [hep-th]\\

\end{flushright}
\title{A new look at the modified Coulomb potential in a strong magnetic field}
\author{N. Sadooghi$^{1,2,}$}\email{sadooghi@physics.sharif.edu}
\author{A. Sodeiri Jalili$^{1}$}
\affiliation{$^{1}$Department of Physics, Sharif University of
Technology, P.O. Box 11365-9161, Tehran-Iran\\
$^{2}$Institute for Studies in Theoretical Physics and Mathematics
(IPM)\\ School of Physics, P.O. Box 19395-5531, Tehran-Iran}
\date{May 30, 2007}
\begin{abstract}
\noindent The static Coulomb potential of Quantum Electrodynamics
(QED) is calculated in the presence of a strong magnetic field in
the lowest Landau level (LLL) approximation using two different
methods. First, the vacuum expectation value of the corresponding
Wilson loop is calculated perturbatively in two different regimes of
dynamical mass $m_{dyn.}$, {\it i.e.}, $|{\mathbf{q}}_{\|}^{2}|\ll
m_{dyn.}^{2}\ll |eB|$ and $m_{dyn.}^{2}\ll
|\mathbf{q}_{\|}^{2}|\ll|eB|$, where $\mathbf{q}_{\|}$ is the
longitudinal components of the momentum relative to the external
magnetic field $B$. The result is then compared with the static
potential arising from Born approximation. Both results coincide.
Although the arising potentials show different behavior in the
aforementioned regimes, a novel dependence on the angle $\theta$
between the particle-antiparticle's axis and the direction of the
magnetic field is observed. In the regime
$|{\mathbf{q}}_{\|}^{2}|\ll m_{dyn.}^{2}\ll |eB|$, for strong enough
magnetic field  and depending on the angle $\theta$, a qualitative
change occurs in the Coulomb-like potential; Whereas for
$\theta=0,\pi$ the potential is repulsive, it exhibits a minimum for
angles $\theta\in]0,\pi[$.
\end{abstract}
 \maketitle
\section{Introduction}
\par\noindent
Chiral symmetry plays an important role in elementary particle
physics. It has been known \cite{new-phases} for some time that QED,
in addition to the familiar weak-coupling phase, may have a
nonperturbative strong coupling phase, characterized by spontaneous
chiral symmetry breaking \cite{leung}. The existence of this new
phase was exploited in a novel interpretation of the multiple
correlated and narrow peak structures in electron and positron
spectra observed at GSI several years ago \cite{electron}. According
to this scenario the electron-positron peaks are due to the decay of
a bound electron-positron system formed in the new QED phase induced
by a strong and rapidly varying electromagnetic field which is
present in the neighborhood of a colliding heavy ions. It is
therefore of great interest to investigate whether background
fields, such as constant magnetic fields, can potentially induce
chiral symmetry breaking in gauge theories and lead eventually to
the formation of chiral symmetry breaking fermion condensate
$\langle\bar{\psi}\psi \rangle$ and a dynamically generated fermion
mass.
\par
Indeed, the magnetic catalysis of dynamical chiral symmetry breaking
has been established as a universal phenomenon in $2+1$ and $3+1$
dimensions \cite{miransky1-5, leung}. According to these results,
even at the weakest attractive interaction between fermions, a
constant magnetic field leads to the generation of a fermion
dynamical mass. The essence of this effect is the dimensional
reduction $D\to D-2$ in the dynamics of fermion pairing in a
magnetic field, which arises from the fact that the motion of
charged particles is restricted in directions perpendicular to the
magnetic field. It is believed that at weak coupling this dynamics
is dominated by the lowest Landau level (LLL). The magnetic
catalysis is not only interesting from purely fundamental point of
view, but it has potential application in condensed matter physics
\cite{cond-matter} and cosmology \cite{cosmology}.
\par
In this paper we are interested on the static potential between the
particle and antiparticle in the presence of a strong constant
magnetic field. The potential produced by a point electric charge
placed into a constant magnetic field is recently calculated in
\cite{shabad-1}. It is shown that the standard Coulomb law is
modified by the vacuum polarization arising in the external magnetic
field. Here, since the vacuum polarization component, taken in
one-loop approximation, grows linearly with the magnetic field, a
scaling regime occurs in the limit of infinite magnetic field. The
scaling regime implies a short range character of interaction,
expressed as Yukawa law
\begin{eqnarray}\label{shabad}
V(\mathbf{x})=-\frac{\alpha e^{-M_{\gamma}R}}{R},
\qquad\qquad\mbox{with}\qquad\qquad
M_{\gamma}\equiv\sqrt{\frac{2\alpha |eB|N_{f}}{\pi}},
\end{eqnarray}
where $R\equiv |\mathbf{x}|$ and $M_{\gamma}$ is the photon mass.
\par
In the present work, we will determine anew the potential $V(x)$
between the particle-antiparticle pair in the presence of a constant
magnetic field in the LLL approximation. Although our method differs
essentially from the analytic methods used in \cite{shabad-1}
leading to (\ref{shabad}), our result is indeed consistent with this
potential.
\par
To determine the fermion-antifermion potential, we will first
compute perturbatively the vacuum expectation value (vev) of a
Wilson loop of a static fermion-antifermion pair for large Euclidean
time.\footnote{Here, to determine the potential in the LLL
approximation, we will use the full photon propagator in the
presence of a strong magnetic field \cite{loskutov} in the LLL
approximation. In \cite{loskutov}, it is shown that the full photon
propagator depends on the dynamical mass $m_{dyn.}$, which is
calculated nonperturbatively in \cite{miransky1-5}. Thus although
our method is a perturbative one, the result depends automatically
on a parameter which is determined nonperturbatively.} In
\cite{kogut-paper}, the same perturbative method is used to
determine the quark-antiquark potential of ordinary QCD. At one-loop
level it is shown to be
\begin{eqnarray}\label{Ax1}
V(R)= -C\frac{e^{2}}{4\pi R}\left(1+\frac{11
e^{2}}{8\pi^{2}}\ln\frac{R}{a}\right),
\end{eqnarray}
where
$C\equiv\frac{1}{4}\mbox{tr}\left(\lambda^{a}\lambda^{a}\right)$ is
the trace over the product of Gell-Mann matrices $\lambda^{a}$, and
$a^{-1}$ is the UV cutoff parameter. Using the same idea, we will
determine the potential of a point charged particle in the external
magnetic field. Eventually, we will compare our result with the
modified Coulomb potential from a semi-classical Born approximation.
In the regime of LLL dominance, we will consider two different
regions of dynamical mass, $ |{\mathbf{q}}_{\|}^{2}|\ll
m_{dyn.}^{2}\ll |eB|$ and $m_{dyn.}^{2}\ll
|\mathbf{q}_{\|}^{2}|\ll|eB|$, separately. Here, $\mathbf{q}_{\|}$
is the longitudinal component of the momentum with respect to the
direction of the external magnetic field. We will show that the
potential in the regime $|{\mathbf{q}}_{\|}^{2}|\ll m_{dyn.}^{2}\ll
|eB|$  has the general form of a modified Coulomb potential
\begin{eqnarray}\label{A4}
V_{1}(R,\theta)=-\frac{\alpha}{
R}\left({\cal{A}}_{1}(\alpha,\theta)-\frac{\gamma{\cal{A}}_{2}(\alpha,\theta)}{R^{2}}+
\frac{\gamma^{2}{\cal{A}}_{3}(\alpha,\theta)}{R^{4}}\right),\qquad\mbox{with}\qquad\gamma\equiv\frac{2\alpha}{3\pi
m_{dyn.}^{2}},
\end{eqnarray}
and in the regime $m_{dyn.}^{2}\ll |{\mathbf{q}}_{\|}^{2}|\ll |eB|$
has the form of a Yukawa-like potential
\begin{eqnarray}\label{A5}
V_{2}(R,\theta)=-\frac{\alpha\ e^{-
M_{\mbox{\tiny{eff.}}}R}}{(1-\frac{\alpha}{\pi})
g(\theta)R},\qquad\qquad \mbox{with}\qquad\qquad
M_{\mbox{\small{eff.}}}(\theta) \equiv
g(\theta)\sqrt{\frac{2\alpha|eB|}{\pi}},
\end{eqnarray}
where ${\cal{A}}_{i},i=1,2,3$ in (\ref{A4}) and $g(\theta)$ in
(\ref{A5}) will be calculated exactly in Sec. III. In (\ref{A4}) as
well as (\ref{A5}), $\theta$ is the angle between the
particle-antiparticle axis and the direction of the magnetic field.
Up to this explicit novel dependence on the angle $\theta$, the
potential $V_{2}(R,\theta)$ from (\ref{A5}) is comparable with the
potential (\ref{shabad}) from \cite{shabad-1}. As a consequence of
this $\theta$-dependence, the effective photon mass
$M_{\mbox{\small{eff.}}}(\theta)$  in (\ref{A5}) is, in contrast to
the photon mass $M_{\gamma}$ in (\ref{shabad}), a function of
$\theta$.
\par
In the regime $|{\mathbf{q}}_{\|}^{2}|\ll m_{dyn.}^{2}\ll |eB|$, it
can be shown that for large enough magnetic field and depending on
the angle $\theta$, a qualitative change occurs in the Coulomb-like
potential $V_{1}(R,\theta)$; Whereas for $\theta=0,\pi$ the
potential is repulsive, it exhibits a minimum for angles
$\theta\in]0,\pi[$ and distances $R\leq 0.005$ fm. The position of
this minimum is proportional to $1/\sqrt{B}$ and the depth of the
potential at $R_{min}$ increases with the magnetic field. The exact
value of the strong magnetic field will be determined in Sec. V. We
interpret the appearance of such a minimum as a possibility for
bound state formation in $D=4$ dimensions. A rigorous proof of the
bound state formation in the above potentials $V_{1}$ and $V_{2}$ is
the subject of a separate investigation and is beyond the scope of
this paper.\footnote{A nonperturbative analysis of the corresponding
Schr\"odinger equation describing the Nambu-Goldstone modes and
arising from a Bethe-Salpeter equation for bound states shows that
at least one bound state can be formed in the attractive potential
in $D=4$ dimensions \cite{miransky1-5} (for more details, see the
explanation in the last paragraph of Sec. II).}
\par
The organization of the paper is as follows. In Sec. II, a brief
review of QED in a strong magnetic field containing some important
results from \cite{miransky1-5} will be presented. In Sec. III, the
static potential of QED in a strong magnetic field will be
calculated perturbatively by determining the vev of the Wilson loop
of a static fermion-antifermion pair in two regimes of the dynamical
mass $|{\mathbf{q}}_{\|}^{2}|\ll m_{dyn.}^{2}\ll |eB|$ and
$m_{dyn.}^{2}\ll |\mathbf{q}_{\|}^{2}|\ll|eB|$ in the LLL. In Sec.
IV a semi-classical Born approximation will be used to determine the
same particle-antiparticle potentials.  In Section V, a qualitative
analysis of these potentials will be performed and the role played
by the angle $\theta$ will be discussed in detail. Sec. VI
summarizes our results.
\section{QED in a strong magnetic field}
\setcounter{equation}{0}\par\noindent In this section we will
briefly review the characteristics of fermions and photons in a
constant external magnetic field. To this purpose, let us start with
the QED Lagrangian density
\begin{eqnarray}\label{B1}
{\cal{L}}=-\frac{1}{4}F_{\mu\nu}F^{\mu\nu}+\bar{\psi}\gamma^{\mu}\left(i\partial_{\mu}+
eA_{\mu}\right)\psi-m\bar{\psi}\psi,
\end{eqnarray}
where the vector field $A_{\mu}=a_{\mu}+A_{\mu}^{ext.}$, where
$a_{\mu}$ is an Abelian quantum gauge field, $F_{\mu\nu}$ is the
corresponding field strength, and $A_{\mu}^{ext.}$ describes an
external electromagnetic field. In this paper we will choose the
symmetric gauge for $A_{\mu}^{ext.}$, {\it i.e.},
\begin{eqnarray}\label{B2}
A_{\mu}^{ext.}=\frac{B}{2}\left(0,x_{2},-x_{1},0\right).
\end{eqnarray}
This leads to a magnetic field in $x_{3}$ direction. From now on,
the longitudinal $(0,3)$ directions will be denoted by
$\mathbf{x}_{\|}$ and the transverse directions $(1,2)$ by
$\mathbf{x}_{\perp}$. Using the Schwinger proper time formalism
\cite{schwinger-1}, it is possible to derive the fermion and photon
propagator in this gauge. As for the fermion propagator, it is given
by
\begin{eqnarray}\label{B3}
{\cal{S}}_{F}(x,y)&=&\exp\left(\frac{ie}{2}\left(x-y\right)^{\mu}A_{\mu}^{ext.}(x+y)\right){S}(x-y)\nonumber\\
&=&\exp\left(\frac{ieB}{2}\epsilon^{ab}x_{a}y_{b}\right)
{S}(x-y),\qquad\qquad a,b=1,2.
\end{eqnarray}
Here, the first factor containing the external gauge field,
$A_{\mu}^{ext.}$, is the Schwinger line integral \cite{schwinger-1}.
The Fourier transform of the translational invariant part ${S}(x-y)$
reads
\begin{eqnarray}\label{B4}
\tilde{S}(k)&=&i\int\limits_{0}^{\infty}ds\ e^{-ism^{2}}\
\exp\left(is\big[k_{\|}^{2}-\frac{k_{\perp}^{2}}{eBs\cot(eBs)}\big]\right)
\nonumber\\
&&\times \bigg\{\left(m+\mathbf{\gamma}^{\|}\cdot
\mathbf{k}_{\|}\right)\left(1-\gamma^{1}\gamma^{2}\tan(eBS)\right)-\mathbf{\gamma}^{\perp}\cdot
\mathbf{k}_{\perp}\left(1+\tan^{2}(eBs)\right)\bigg\},
\end{eqnarray}
where $\mathbf{k}_{\|}=(k_{0},k_{3})$ and
$\mathbf{\gamma}_{\|}=(\gamma_{0},\gamma_{3})$ and
$\mathbf{k}_{\perp}=(k_{1},k_{2})$ and
$\mathbf{\gamma}_{\perp}=(\gamma_{1},\gamma_{2})$.  After performing
the integral over $s$,  $\tilde{S}(k)$ can be decomposed as follows
\begin{eqnarray}\label{B5}
\tilde{S}(k)=ie^{-\frac{k_{\perp}^{2}}{|eB|}}\sum\limits_{n=0}^{\infty}(-1)^{n}\
\frac{D_{n}(eB,k)}{k_{\|}^{2}-m^{2}-2|eB|n},
\end{eqnarray}
with $D_{n}(eB,k)$ expressed through the generalized Laguerre
polynomials $L_{m}^{\alpha}$
\begin{eqnarray}\label{B6}
D_{n}(eB,k)=(\gamma^{\|}\cdot \mathbf{k}_{\|}+m)\ \bigg\{ 2\
{\cal{O}}\bigg[L_{n}\left(2\rho\right)-
L_{n-1}\left(2\rho\right)\bigg]+4\gamma^{\perp}\cdot
k_{\perp}L_{n-1}^{1}\left(2\rho\right) \bigg\}.
\end{eqnarray}
Here, we have introduced $\rho\equiv \frac{k_{\perp}^{2}}{|eB|}$ and
\begin{eqnarray}\label{B7}
{\cal{O}}\equiv\frac{1}{2}\left(1-i\gamma^{1}\gamma^{2}\mbox{sign}(eB)\right).
\end{eqnarray}
Relation (\ref{B5}) suggests that in the IR region, with
$|\mathbf{k}_{\|}|, |\mathbf{k}_{\perp}|\ll \sqrt{|eB|}$, all the
higher Landau levels with $n\geq 1$ decouple and only the lowest
Landau level (LLL) with $n=0$ is relevant. In strong magnetic field
limit, the full fermion propagator (\ref{B3}) can therefore be
decomposed into two independent transverse and longitudinal parts
\cite{miransky1-5}
\begin{nedalph}\label{B8a}
{\cal{S}}_{F}(x,y)=S_{\|}(\mathbf{x}_{\|}-\mathbf{y}_{\|})P(\mathbf{x}_{\perp},\mathbf{y}_{\perp})\
,
\end{eqnarray}
with the longitudinal part
\begin{eqnarray}\label{B8b}
S_{\|}(\mathbf{x}_{\|}-\mathbf{y}_{\|})=\int
\frac{d^{2}k_{\|}}{(2\pi)^{2}}\
e^{i\mathbf{k}_{\|}\cdot(\mathbf{x}-\mathbf{y})^{\|}}\
\frac{i{\cal{O}}}{\gamma^{\|}\cdot \mathbf{k}_{\|}-m},
\end{eqnarray}
and the transverse part
\begin{eqnarray}\label{B8c}
P(\mathbf{x}_{\perp},\mathbf{y}_{\perp})=\frac{|eB|}{2\pi}\exp\left(\frac{ieB}{2}\epsilon^{ab}x^{a}y^{b}-
\frac{|eB|}{4}\left(\mathbf{x}_{\perp}-\mathbf{y}_{\perp}\right)^{2}\right),\qquad
a,b=1,2.
\end{nedalph}

\par\noindent
The photon propagator ${\cal{D}}_{\mu\nu}$ of QED in an external
constant magnetic field in one-loop approximation with fermions from
lowest Landau level (LLL) is calculated explicitly in
\cite{loskutov, miransky1-5}. It is given by
\begin{eqnarray}\label{B9}
i{\cal{D}}_{\mu\nu}(q)=\frac{g_{\mu\nu}^{\perp}}{q^{2}}+\frac{q_{\mu}^{\|}q_{\nu}^{\|}}{q^{2}q_{\|}^{2}}+
\frac{\left(g_{\mu\nu}^{\|}-q_{\mu}^{\|}q_{\nu}^{\|}/q_{\|}^{2}\right)}{q^{2}+q_{\|}^{2}
\Pi(q_{\perp}^{2},q_{\|}^{2})}-{\xi}\
\frac{q_{\mu}q_{\nu}}{(q^{2})^{2}},
\end{eqnarray}
where $\xi$ is an arbitrary gauge parameter. Since the LLL fermions
couple only to the longitudinal $(0,3)$ components of the photon
fields, no polarization effects are present in the transverse
$(1,2)$ components of ${\cal{D}}_{\mu\nu}(q)$. Therefore, the full
photon propagator in the LLL approximation is given by the
Feynman-like covariant propagator \cite{miransky1-5}
\begin{eqnarray}\label{B10}
i\widetilde{{\cal{D}}}_{\mu\nu}(q)=\frac{g_{\mu\nu}^{\|}}{q^{2}+\mathbf{q}_{\|}^{2}\Pi\left(\mathbf{q}_{\|}^{2},\mathbf{q}_{\perp}^{2}\right)},
\end{eqnarray}
with $\Pi(q_{\perp}^{2},q_{\|}^{2})$ having the form
\cite{kuznetsov}
\begin{eqnarray}\label{B10-a}
\Pi(q_{\perp}^{2},q_{\|}^{2})=\frac{2\alpha|eB|N_{f}}{\mathbf{q}_{\|}^{2}}e^{-\frac{q_{\perp}^{2}}{2|eB|}}H\left(\frac{\mathbf{q}^{2}_{\|}}
{4m^{2}_{dyn.}}\right).
\end{eqnarray}
Here $N_{f}$ is the number of fermion flavors and $\alpha\equiv
\frac{e^{2}}{4\pi}$, where $e$ is the running coupling. Further,
$H(z)$ in (\ref{B10-a}) is defined by
\begin{eqnarray}\label{B10-b}
H(z)\equiv
\frac{1}{2\sqrt{z(z-1)}}\ln\left(\frac{\sqrt{1-z}+\sqrt{-z}}{\sqrt{1-z}-\sqrt{-z}}\right)-1.
\end{eqnarray}
Expanding this expression for $|{\mathbf{q}}_{\|}^{2}|\ll
m_{dyn.}^{2}\ll |eB|$ and $m_{dyn.}^{2}\ll
|\mathbf{q}_{\|}^{2}|\ll|eB|$ leads to
\begin{eqnarray}
\Pi({\mathbf{q}}_{\perp}^{2},{\mathbf{q}}_{\|}^{2})\simeq\
+\frac{\alpha|eB|N_{f}}{3\pi
m_{dyn.}^{2}}e^{-\frac{q_{\perp}^{2}}{2|eB|}}&\qquad\mbox{for}\qquad&
|\mathbf{q}_{\|}^{2}|\ll m_{dyn.}^{2}\ll|eB|,\label{B11}\\
\Pi({\mathbf{q}}_{\perp}^{2},{\mathbf{q}}_{\|}^{2})\simeq
-\frac{2\alpha|eB|N_{f}}{\pi\
\mathbf{q}_{\|}^{2}}e^{-\frac{q_{\perp}^{2}}{2|eB|}}&\qquad\mbox{for}\qquad&m_{dyn.}^{2}\ll
|\mathbf{q}_{\|}^{2}|\ll|eB|.\label{B12}
\end{eqnarray}
In \cite{miransky1-5}, it is shown that the kinematic region mostly
responsible for generating the fermion mass is with the dynamical
mass, $m_{dyn.}$, satisfying  $m_{dyn.}^{2}\ll
|\mathbf{q}_{\|}^{2}|\ll|eB|$. Plugging (\ref{B12}) in the full
photon propagator (\ref{B10}) and assuming that
$|\mathbf{q}_{\perp}^{2}|\ll |eB|$, we get
\begin{eqnarray}\label{LL2}
\widetilde{\cal{D}}_{\mu\nu}(q)\approx
-\frac{ig_{\mu\nu}^{\|}}{q^{2}-M_{\gamma}^{2}},\qquad
\mbox{with}\qquad M_{\gamma}^{2}=\frac{2\alpha|eB|N_{f}}{\pi}.
\end{eqnarray}
The appearance of a finite photon mass $M_{\gamma}$ is the result of
the dimensional reduction $3+1\to 1+1$ in the presence of a constant
magnetic field. This phenomenon can be understood as a reminiscent
of the Higgs effect in the $1+1$ dimensional Schwinger model, where
the photon acquires also a finite mass.
\par
As for the dynamically generated fermion mass $m_{dyn.}$, it can be
determined by solving the corresponding Schwinger-Dyson (SD)
equation in the rainbow (ladder) approximation, where the effects of
dynamical fermions are neglected. In this approximation, the gauge
invariant dynamical mass is shown to have the form
\cite{miransky1-5}
\begin{eqnarray}\label{B13}
m_{dyn.}=C\sqrt{eB}\exp\left(-\frac{\pi}{2}\left(\frac{\pi}{2\alpha}\right)^{1/2}\right),
\end{eqnarray}
where the constant $C$ is of order one.\footnote{The problem of
gauge invariance of chiral symmetry breaking induced by the magnetic
field found in the ladder QED is investigated in \cite{ferrer-WI}.}
In the improved rainbow approximation, however, the expression for
$m_{dyn.}$ takes the following form \cite{miransky1-5}
\begin{eqnarray}\label{B14}
m_{dyn.}=\tilde{C}\sqrt{|eB|}F(\alpha)\exp\left(-\frac{\pi}{\alpha\ln\left(C_{1}/\alpha
N_{f}\right)}\right),
\end{eqnarray}
where $F(\alpha)\simeq (N_{f}\alpha)^{1/3}$, $C_{1}\simeq 1.82\pm
0.06$ and $\tilde{C}\sim O(1)$.
\par
The dynamical mass, $m_{dyn.}$, plays the role of energy in the
two-dimensional Schr\"odinger equation
\begin{eqnarray}\label{SRE}
\left(-\Delta+m_{dyn.}^{2}+V(\mathbf{x})\right)\Psi(\mathbf{x})=0,
\qquad\qquad \mbox{with}\qquad\qquad\Delta\equiv
\frac{\partial^{2}}{\partial x_{3}^{2}}+\frac{\partial^{2}}{\partial
x_{4}^{2}}.
\end{eqnarray}
Here, the attractive potential has the form \cite{miransky1-5}
\begin{eqnarray}\label{B15}
V(\mathbf{x})=\frac{\alpha |eB|}{\pi}\
\exp\left(\frac{R^{2}|eB|}{2}\right)\mbox{Ei}\left(-\frac{R^{2}|eB|}{2}\right),
\end{eqnarray}
and behaves as
\begin{eqnarray}\label{B16}
V(\mathbf{x})&\simeq&-\frac{2\alpha}{\pi}\frac{1}{R^{2}},\
\qquad\qquad\qquad\qquad\qquad
R\to\infty,\nonumber\\
V(\mathbf{x})&\simeq&-\frac{\alpha
|eB|}{\pi}\left(\gamma_{E}+\ln\frac{2}{R^{2}|eB|}\right),\qquad R\to
0,
\end{eqnarray}
where $\gamma_{E}\simeq 0.577$ is the Euler constant. To determine
the long and short range behavior of the potential, the asymptotic
behavior of $\mbox{Ei}(x)$ is used \cite{gradshteyn}. In this
context, the problem concerning bound state formation is reduced to
finding the spectrum of bound states of the Schr\"odinger equation
(\ref{SRE}) with the attractive potential (\ref{B15}). In
\cite{miransky1-5}, it is shown that in $D=4$ dimensions at least
one bound state can be formed. Figure 1 shows the behavior of
$V(\mathbf{x})$ for different $B=10^3, 10^4, 10^5, 10^9$ and
$B=10^{12}$ (from left to right) as a function of $R$.\footnote{Note
that discussion here is purely qualitative. It is nevertheless
possible to give the strength of the magnetic fields in Gauss
through the relation  $ 6.8\times 10^{19}B\simeq H\ \mbox{in
Gau\ss}$ \cite{ng, ferrer}.  Using this relation the potential
$V(R)$ is given in GeV=$10^9$ eV and the distance $R$ in
fm=$10^{-15}$ m.} According to this result, whereas for $B< 10^9$ at
each given distance the absolute value of the potential increases
with increasing the magnetic field, the shape of the potential does
not change for $B\geq 10^9$.
\begin{figure}[b]
\includegraphics[width=9cm, height=5.5cm]{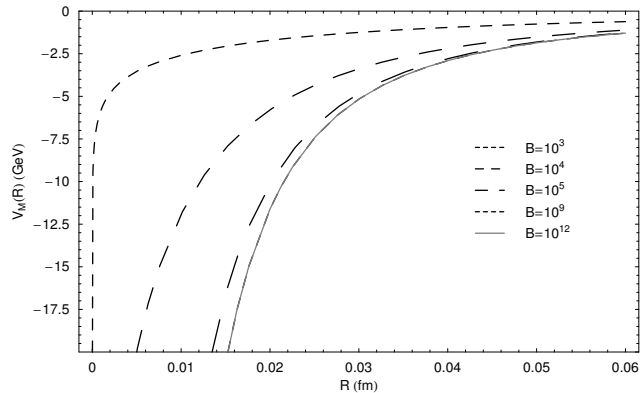}
\caption{Potential $V(\mathbf{x})$ from (\ref{B15}) of a
particle-antiparticle pair for different magnetic fields. According
to this result whereas for $B\leq 10^9$ at each given distance the
absolute value of the potential increases with increasing the
magnetic field, the shape of the potential does not change for
$B\geq 10^9$.}
\end{figure}
\par\noindent
In the next section we will determine the static Coulomb potential
between a fermion-antifermion pair in two aforementioned regimes in
the LLL approximation by calculating the vev of the Wilson loop
perturbatively. Note that although our method is a perturbative one,
our results depend explicitly on $m_{dyn.}$ which is determined
nonperturbatively in the literature \cite{miransky1-5}.
\section{The Wilson loop and the modified Coulomb  potential in a strong magnetic field}
\setcounter{equation}{0}\par\noindent In ordinary Quantum Field
Theory with no background magnetic field, the Wilson loop appears as
one of the most efficient tools for probing the large distance
properties of strong coupling QCD. It provides a natural criterion
for confinement through the area law. In contrast to QED, where the
field lines connecting a pair of opposite charges are allowed to
spread, one expects that in QCD the quarks within a hadron are the
sources of a chromoelectric flux which is concentrated within narrow
tubes connecting the constituents. Since the energy is not allowed
to spread, the potential of a quark-antiquark pair
$\langle\bar{q}q\rangle$ will increase with their separation, as
long as vacuum polarization effect do not screen their color charge.
This picture of confinement can be checked by computing the
nonperturbative potential between a static quark-antiquark pair in
the path integral formalism.
\par
It is the main purpose of this section to investigate the properties
of the Wilson loop concerning the bound state problem of QED in the
presence of a strong magnetic field in the LLL approximation. Here,
the magnetic field plays the role of a strong catalyst and even the
weakest attractive potential between fermions is enough for
dynamical mass generation and bound state formation. Before
calculating the modified Coulomb potential by determining the vev of
the Wilson line of a static fermion-antifermion pair in two regimes
of dynamical mass, $ |{\mathbf{q}}_{\|}^{2}|\ll m_{dyn.}^{2}\ll
|eB|$ and $m_{dyn.}^{2}\ll |\mathbf{q}_{\|}^{2}|\ll|eB|$, we briefly
review the path integral formulation leading to the Coulomb
potential between slowly moving particles in ordinary QED. Here, we
keep closely on the notations of \cite{rothe} and the classical
review article \cite{kogut-paper}.
\par
Imagine creating a fermion-antifermion pair, $\bar{\psi}\psi$, at
space time point $x_{\mu}=0$ and adiabatically separate them to a
relative distance $R$. This configuration will be hold for a time
$T\to \infty$. Finally we bring $\bar{\psi}\psi$ pair back together
and let them annihilate. The Euclidean amplitude for this process is
the matrix element of the Hamiltonian evolution operator $e^{-HT}$
between the initial and final states, $|i\rangle$ and $|f\rangle$,
respectively
\begin{eqnarray}\label{C1}
\langle i|e^{-HT}|f\rangle.
\end{eqnarray}
Here, $|i\rangle$ and $|f\rangle$ represents a $\bar{\psi}\psi$ pair
a distance $R$ apart and $H$ is the Hamiltonian (for more details
concerning the exact mathematical structure of the initial and final
states, $|i\rangle$ and $|f\rangle$, see \cite{rothe}). In the path
integral representation the matrix element (\ref{C1}) can be
expressed by
\begin{eqnarray}\label{C2}
\langle i|e^{-HT}|f\rangle=\frac{\int
{\cal{D}}A_{\mu}{\cal{D}}\psi{\cal{D}}\bar{\psi}\ e^{-S+ie\int
A_{\mu}(x)j_{\mu}(x)
d^{4}x}}{\int{\cal{D}}A_{\mu}{\cal{D}}\psi{\cal{D}}\bar{\psi}\
e^{-S}},
\end{eqnarray}
where we have already skipped the subscript $E$ for the Euclidean
action $S$. Here, the current $j_{\mu}$ should describe the closed
worldline ${{C}}$ of the creation and annihilation of
$\bar{\psi}\psi$ pair. For a closed worldline of the heavy fermions
(\ref{C2}) simplifies to \cite{rothe}
\begin{eqnarray}\label{C2-a}
\langle i|e^{-HT}|f\rangle=\frac{\int
{\cal{D}}A_{\mu}{\cal{D}}\psi{\cal{D}}\bar{\psi}\
e^{-S+ie\oint\limits_{{{C}}} A_{\mu}(x)
dx_{\mu}}}{\int{\cal{D}}A_{\mu}{\cal{D}}\psi{\cal{D}}\bar{\psi}\
e^{-S}}.
\end{eqnarray}
Since in a closed path $|i\rangle$ and $|f\rangle$ are identical and
since the process is static, the Hamiltonian $H$ is purely potential
and the left hand side (l.h.s) of (\ref{C2-a}) is
\begin{eqnarray}\label{C3}
\langle i|e^{-HT}|f\rangle=e^{-V(R)T}.
\end{eqnarray}
Here, $V(R)$ is the fermion-antifermion potential. Taking the
logarithm of (\ref{C2-a}) by plugging (\ref{C3}) on its l.h.s, it is
given by
\begin{eqnarray}\label{C4}
V(R)=-\lim\limits_{T\to \infty} \frac{1}{T}\ln\langle W_{C}[A]
\rangle,
\end{eqnarray}\label{C5}
where we have introduced the ``loop-correlation'' function -- the
{\it Wilson loop} \cite{wegner, wilson, polyakov},
\begin{eqnarray}\label{C5-a}
W_{C}[A]\equiv e^{ie\oint\limits_{C} A_{\mu}(x) dx_{\mu}},
\end{eqnarray}
and used the definition
\begin{eqnarray}\label{C6}
\langle {\cal{O}} \rangle\equiv \frac{\int
{\cal{D}}A_{\mu}{\cal{D}}\psi{\cal{D}}\bar{\psi}\ {\cal{O}}\
e^{-S}}{\int{\cal{D}}A_{\mu} {\cal{D}}\psi{\cal{D}}\bar{\psi}\
e^{-S}}.
\end{eqnarray}
For $S$ being the QED action, the Coulomb potential
$V(R)=-\frac{e^{2}}{4\pi R}$ can be analytically calculated in the
quenched approximation, {\it i.e.} when the vacuum polarization
effects arising from the presence of dynamical fermions are
neglected. To calculate this Coulomb potential, one expands $\langle
W_{C}[A] \rangle$ in powers of the background field $A_{\mu}$, to
get
\begin{eqnarray}\label{C7}
\langle W_{C}[A] \rangle=\langle 1+ie\oint_{C}
dx_{\mu}A_{\mu}(x)-\frac{e^{2}}{2}\oint_{C} \oint_{C}
dx_{\mu}dy_{\nu} A_{\mu}(x)A_{\nu}(y)+\cdots\rangle.
\end{eqnarray}
Here, the second term including only one gauge field does not
contribute. As for the other terms, the terms with an odd number of
external photon lines does not contribute to the above expansion.
This is because of the Furry's theorem, that holds in ordinary QED
in contrary to QED in the presence of external magnetic field
\footnote{The phenomenon of photon splitting $\gamma\to 2\gamma$ in
the presence of external magnetic field \cite{adler} is the best
counterexample for the validity of Furry's theorem for QED in
external magnetic field.}. Plugging (\ref{C7}) in (\ref{C4}), the
Coulomb potential is given by
\begin{eqnarray}\label{C8}
V(R)&=&\lim\limits_{T\to\infty}-\frac{1}{T}\ln\left(1-\frac{e^{2}}{2}\oint_{C}\oint_{C}
dx_{\mu}dy_{\nu}\ D_{\mu\nu}(x,y)+\cdots\right)\nonumber\\
&=&\lim\limits_{T\to\infty}\frac{e^{2}}{2T}\oint_{C}\oint_{C}
dx_{\mu}dy_{\nu}\ D_{\mu\nu}(x,y)+{\cal{O}}(e^{3}),
\end{eqnarray}
where we have expanded the $\ln(1-\cdots)$ for weak coupling
constant $e$. The integrand is the photon propagator in the
coordinate space
\begin{eqnarray}\label{C9}
D_{\mu\nu}(x,y)\equiv \langle
A_{\mu}(x)A_{\nu}(y)\rangle=\frac{\delta_{\mu\nu}}{4\pi^{2}(x-y)^{2}}.
\end{eqnarray}
To compute the double integral (\ref{C8}), one chooses a rectangular
contour. Because, in the Euclidean space, the integrand (\ref{C9})
is proportional to $\delta_{\mu\nu}$, the double integral receives a
contribution when $x$ and $y$ are located on the segments of
integration contour which are parallel to each other. The double
integral therefore reduces to
\begin{eqnarray}\label{C9-a}
V(R)&=&\lim\limits_{T\to\infty}\frac{-e^{2}}{T}\int_{0}^{T}dT_{1}\int_{0}^{T}dT_{2}\
D_{00}(R,T_{1}-T_{2})\nonumber\\
&=&\lim\limits_{T\to\infty}\frac{-e^{2}}{T}\int_{0}^{T}dT_{1}\int_{0}^{T}dT_{2}\
\frac{1}{4\pi^{2}\big[R^{2}+\left(T_{1}-T_{2}\right)^{2}\big]}.
\end{eqnarray}
Here $R\equiv|\mathbf{x}-\mathbf{y}|$. Performing now the double
integration, one arrives first at \cite{rothe}
\begin{eqnarray}\label{C10}
V(R)=\lim\limits_{T\to\infty}-\frac{e^{2}}{2\pi^{2}}\left(\frac{1}{R}\arctan\frac{T}{R}+\frac{1}{2T}\ln\left(1+\frac{T^{2}}{R^{2}}\right)\right),
\end{eqnarray}
and then after taking the limit $T\to\infty$, at the Coulomb
potential of a fermion-antifermion pair
\begin{eqnarray}\label{C11}
V(R)=-\frac{\alpha}{R}.
\end{eqnarray}
After this brief excursion, let us now turn back to our original
problem of computing the modified Coulomb potential of QED in the
presence of external strong magnetic field. In the regime of lowest
Landau level (LLL), the full photon propagator
$\widetilde{{\cal{D}}}_{\mu\nu}$ in the momentum space is given by
(\ref{B10}), where $\Pi(\mathbf{q}_{\|}^{2},\mathbf{q}_{\perp}^{2})$
is defined in (\ref{B10-a}). It is indeed a very difficult task to
calculate the potential of static fermion-antifermion pair for full
photon propagator (\ref{B10}). We will therefore consider only two
different regimes of dynamical mass $|{\mathbf{q}}_{\|}^{2}|\ll
m_{dyn.}^{2}\ll |eB|$ and $m_{dyn.}^{2}\ll
|\mathbf{q}_{\|}^{2}|\ll|eB|$, with
$\Pi(\mathbf{q}_{\|}^{2},\mathbf{q}_{\perp}^{2})$ given in
(\ref{B11}) and (\ref{B12}), respectively, and will calculate the
modified Coulomb potential $V(x)$ in these two regimes.
\subsection{Modified Coulomb potential in $ |{\mathbf{q}}_{\|}^{2}|\ll m_{dyn.}^{2}\ll |eB|$ regime}
\par\noindent
Substituting (\ref{B11}) with $N_{f}=1$ in (\ref{B10}) the full
photon propagator in this regimes reads
\begin{eqnarray}\label{C12}
\widetilde{\cal{D}}_{\mu\nu}(q)=-i\frac{g_{\mu\nu}^{\|}}{q^{2}+\frac{\alpha
|eB|}{3\pi
m_{dyn.}^{2}}\mathbf{q}_{\|}^{2}\exp\left(-\frac{\mathbf{q}_{\perp}^{2}}{2|eB|}\right)}.
\end{eqnarray}
To calculate the Coulomb potential by (\ref{C8}), we need the LLL
photon propagator in the coordinate space, {\it i.e.} the Fourier
transformed of (\ref{C12})
\begin{eqnarray}\label{C12-a}
\widetilde{\cal{D}}_{\mu\nu}(x)=-ig_{\mu\nu}^{\|}\int\frac{d^{4}q}{(2\pi)^{4}}\frac{e^{iqx}}{q^{2}+
\frac{\alpha |eB|}{3\pi
m_{dyn.}^{2}}\mathbf{q}_{\|}^{2}\exp\left(-\frac{\mathbf{q}_{\perp}^{2}}{2|eB|}\right)}.
\end{eqnarray}
After a lengthy but straightforward calculation of the above
integral over $q$ (see Appendix A for more details), we arrive at
the LLL photon propagator in the coordinate space
\begin{eqnarray}\label{C13}
\lefteqn{\widetilde{\cal{D}}_{\mu\nu}(R,\theta,T)=
\frac{\delta_{\mu\nu}^{\|}}{4\pi^{2}a_{1}}\bigg[\left(1+\frac{\gamma}{
a_{1}}-\frac{4a_{2}+\gamma R^{2}\sin^{2}\theta}{2\beta
a_{1}^{2}}+\frac{3a_{2}R^{2}\sin^{2}\theta}{2\beta^{2}a_{1}^{3}}\right)
}\nonumber\\
&&+\frac{4\gamma^{2}}{a_{1}^{2}}\left(1-\frac{3R^2\sin^{2}\theta}{2\beta
a_{1}}+\frac{3R^{4}\sin^{4}\theta}{8\beta^{2}a_{1}^{2}}\right)\nonumber\\
 && -\frac{12\gamma
a_{2}}{\beta a_{1}^{3}}\left(2-\frac{2}{\beta
a_{1}}\left(2R^{2}\sin^{2}\theta+\frac{a_{2}}{\gamma}\right)+\frac{5}{4\beta^{2}a_{1}^{2}}\left({R^{4}\sin^{4}\theta}+
\frac{4a_{2}R^{2}\sin^{2}\theta}{\gamma}\right)-\frac{15a_{2}R^{4}\sin^{4}\theta}{8\gamma\beta^{3}a_{1}^{3}}\right)
\nonumber\\
&&-\frac{\gamma}{|eB|\beta a_{1}^{2}}\left(1-\frac{3}{\beta
a_{1}}\left(\frac{R^{2}\sin^{2}\theta}{2}+\frac{a_{2}}{\gamma}\right)+\frac{6}{\beta^{2}a_{1}^{2}}\left(\frac{R^{4}\sin^{4}\theta}{16}+\frac{a_{2}R^{2}\sin^{2}\theta}{\gamma}
\right)-\frac{15a_{2}R^{4}\sin^{4}\theta}{8\gamma
\beta^{3}a_{1}^{3}}\right) \bigg]
.\nonumber\\
\end{eqnarray}
Here,
\begin{eqnarray}\label{C13-a}
\beta^{-1}\equiv {4\left(1+\frac{\alpha|eB|}{3\pi
m_{dyn.}^{2}}\right)}\qquad\mbox{and}\qquad\gamma(\alpha)\equiv
\frac{2\alpha}{3\pi m_{dyn.}^{2}},
\end{eqnarray}
are constant c-numbers, and $a_{i}=a_{i}\left(R,\theta,T\right),\
i=1,2$ are defined by
\begin{eqnarray}\label{C14}
a_{1}(R,\theta,T)&\equiv&T^{2}+ R^{2}
f^{2}(\alpha,\theta),\qquad\mbox{with}\qquad
f^{2}(\alpha,\theta)\equiv 1+\frac{\gamma |eB|}{2}\sin^{2}\theta,\nonumber\\
a_{2}(R,\theta,T)&\equiv&\beta\gamma\left(T^{2}+R^{2}\cos^{2}\theta\right).
\end{eqnarray}
In all the above expressions $R\equiv |\mathbf{x}|$ is the distance
between the static fermion and antifermion pair, $\theta$ is the
angle between the particle-antiparticle axis and the direction of
the magnetic field $B$, {\it i.e.} the $x_{3}$ direction, and
$T\equiv ix_{0}$ is the Euclidean time. Note that here, to determine
the LLL photon propagator, we have to use of the approximation
${\mathbf{q}}_{\perp}^{2}\ll |eB|$, which is valid in the regime of
strong magnetic field.
\par
To determine the static potential between the $\bar{\psi}\psi$ pair,
we have to calculate the double integral [see (\ref{C9-a}) leading
to the expression of the ordinary Coulomb potential]
\begin{eqnarray}\label{C17}
V(R,\theta)=\lim\limits_{T\to \infty}
-\frac{e^{2}}{T}\int_{0}^{T}dT_{1}\int_{0}^{T}dT_{2}\
\widetilde{\cal{D}}_{00}(R,\theta, T_{1}-T_{2}),
\end{eqnarray}
which can be easily simplified to
\begin{eqnarray}\label{C18}
V(R,\theta)= -2e^{2}\int_{0}^{\infty} dT\
\widetilde{\cal{D}}_{00}(R,\theta, T).
\end{eqnarray}
Using now the definition of
$\widetilde{\cal{D}}_{\mu\nu}(R,\theta,T)$ in (\ref{C13}), and
performing the integration over $T$, the modified Coulomb potential
in the regime $ |{\mathbf{q}}_{\|}^{2}|\ll m_{dyn.}^{2}\ll |eB|$ can
be given by
\begin{eqnarray}\label{C19}
V(R,\theta)=-\frac{\alpha}{
R}\left({\cal{A}}_{1}(\alpha,\theta)-\frac{\gamma{\cal{A}}_{2}(\alpha,\theta)}{R^{2}}+
\frac{\gamma^{2}{\cal{A}}_{3}(\alpha,\theta)}{R^{4}}\right),
\end{eqnarray}
with
\begin{eqnarray}\label{C20}
\lefteqn{
{\cal{A}}_{1}(\alpha,\theta)\equiv\frac{1}{f},
}\nonumber\\
\lefteqn{
{\cal{A}}_{2}(\alpha,\theta)\equiv -\frac{1}{4
f^{3}}\left(1-\frac{3
\cos^{2}\theta}{f^{2}}-\frac{3\sin^{2}\theta}{8\beta
f^{2}}+\frac{15\sin^{2}\theta\cos^{2}\theta}{8\beta
f^{4}}\right),}\nonumber\\
\lefteqn{ {\cal{A}}_{3}(\alpha,\theta)\equiv
+\frac{9}{16f^{5}}\left(1-\frac{5\sin^2\theta}{4\beta
f^2}+\frac{35\sin^{4}\theta}{128\beta^{2}f^{4}}\right)
}\nonumber\\
&&-\frac{15\cos^2\theta}{8f^7}\left(3-\frac{7}{4\beta
f^2}\left(2\beta\cos^2\theta+3\sin^2\theta\right)+\frac{63\sin^2\theta}{8\beta^2f^4}
\left(\beta\cos^2\theta+\frac{3\sin^2\theta}{16}\right)-\frac{693\cos^2\theta\sin^4\theta}{256\beta^2 f^6}\right)\nonumber\\
&&-\frac{3}{16|eB|f^{5}\gamma\beta}\left(1-\frac{5}{4\beta
f^{2}}\left(4\beta\cos^{2}\theta+\sin^{2}\theta\right)+\frac{35\sin^{2}\theta}{4\beta^2f^4}\left(\beta\cos^{2}\theta+\frac{\sin^2\theta}{32}\right)-\frac{315\cos^{2}\theta\sin^{4}\theta}{128\beta^2
f^6}\right)
.\nonumber\\
\end{eqnarray}
In Sec. V we will present a qualitative analysis of the above
potential emphasizing the role played by the angle $\theta$ in a
possible bound state formation.
\subsection{Modified Coulomb potential in $m_{dyn.}^{2}\ll |\mathbf{q}_{\|}^{2}|\ll|eB|$ regime}
\par\noindent
To compute the inter-particle potential in this regime, we
substitute (\ref{B12}) in (\ref{B10}). For $N_{f}=1$, the full
photon propagator in the coordinate space reads
\begin{eqnarray}\label{C21}
\widetilde{\cal{D}}_{\mu\nu}(x)=-ig_{\mu\nu}^{\|}\int\frac{d^{4}q}{(2\pi)^{4}}\frac{e^{iqx}}{q^{2}-
\frac{2\alpha|eB|}{\pi}\exp\left(-\frac{\mathbf{q}_{\perp}^{2}}{2|eB|}\right)}.
\end{eqnarray}
The integration over $q$ can be performed using the approximation
$\mathbf{q}_{\perp}^{2}\ll|eB|$ which is valid in the regime of LLL
dominance. After a straightforward calculation (see Appendix B for
more details), the propagator is given by
\begin{eqnarray}\label{C22}
\widetilde{\cal{D}}_{\mu\nu}(R,\theta,T)=\frac{\delta_{\mu\nu}^{\|}}{4\pi^{2}\left(1-\frac{\alpha}{\pi}\right)}
\frac{\zeta}{\sqrt{T^{2}+R^{2}g^{2}(\theta)}}K_{1}\left(\zeta\sqrt{T^{2}+R^{2}g^{2}(\theta)}\right),
\end{eqnarray}
where
\begin{eqnarray}\label{C23}
\zeta\equiv \sqrt{\frac{2\alpha|eB|}{\pi}},\qquad\mbox{and}\qquad
g^2(\theta)\equiv
\cos^{2}\theta+\frac{\sin^{2}\theta}{1-\frac{\alpha}{\pi}}.
\end{eqnarray}
The fermion-antifermion potential in this regime is then calculated
using the expression (\ref{C17}) or equivalently (\ref{C18}) and
reads
\begin{eqnarray}\label{C24}
V(R,\theta)=-\frac{e^{2}\zeta}{2\pi^{2}\left(1-\frac{\alpha}{\pi}\right)}\int_{0}^{\infty}dT\
\frac{1}{\sqrt{T^{2}+R^{2}g^{2}(\theta)}}K_{1}\left(\sqrt{T^{2}+R^{2}g^{2}(\theta)}\right).
\end{eqnarray}
To evaluate the integral we make use of
\begin{eqnarray}\label{C25}
\int_{0}^{\infty} d\tau
\frac{\tau^{2\mu+1}}{\sqrt{(\tau^{2}+z^{2})^{\nu}}}K_{\nu}\left(\zeta\sqrt{\tau^{2}+z^{2}}\right)=
\frac{2^{\mu}\Gamma(\mu+1)}{\zeta^{\mu+1}z^{\nu-\mu-1}}K_{\nu-\mu-1}(\zeta
z),\qquad \zeta>1,\ \mbox{Re}(\mu)>-1.
\end{eqnarray}
Choosing $\tau=T$, $z=Rg(\theta)$, $\nu=1$ and $\mu=-\frac{1}{2}$,
we arrive at
\begin{eqnarray}\label{C26}
V(R,\theta)=-\frac{\alpha}{(1-\frac{\alpha}{\pi})g(\theta)R}e^{-\zeta
g(\theta)R}.
\end{eqnarray}
Here we have used
$K_{\frac{1}{2}}\left(x\right)=\sqrt{\frac{\pi}{2x}}e^{-x}$ and
$\Gamma(\frac{1}{2})=\sqrt{\pi}$. Apart from its $\theta$ dependence
our result is comparable with (\ref{shabad}) from \cite{shabad-1}.
\par
The potential (\ref{C26}) is indeed comparable with the ordinary
attractive Yukawa potential
\begin{eqnarray}\label{C28}
V_{\mbox{\tiny{Yukawa}}}(R)=-\frac{\alpha}{R}e^{-m R},
\end{eqnarray}
where $\alpha\to
\frac{\alpha}{\left(1-\frac{\alpha}{\pi}\right)g(\theta)}$ and the
effective photon mass $m\to
M_{\mbox{\tiny{eff.}}}(\theta)\equiv\zeta g(\theta)$. This result is
also in agreement with the general result about the massive photons
in a strong magnetic fields in the LLL approximation.  As we have
seen in Sec. II, in the regime $m_{dyn.}^{2}\ll
|\mathbf{q}_{\|}^{2}|\ll|eB|$, the $3+1$ dimensional QED in the LLL
approximation is reduced to a $1+1$ dimensional Schwinger model
where the photon acquires a finite mass
$M_{\gamma}=\sqrt{\frac{2\alpha|eB|N_{f}}{\pi}}$, with $N_{f}$ the
number of flavors [see (\ref{LL2}) for more details]. For $N_{f}=1$,
we have $M_{\gamma}=\zeta$. Thus the effective photon mass in this
regime is given by $M_{\mbox{\tiny{eff.}}}=M_{\gamma}g(\theta)$ and
depends explicitly on the angle $\theta$ between the particle axis
and the direction of the external magnetic field. Figure 2 shows the
$\theta$-dependence of $g(\theta)$ which is understood to be the
ratio $\frac{M_{\mbox{\tiny{eff.}}}}{M_{\gamma}}$. For all values of
$\theta\in [0,\pi]$ the effective mass $M_{\mbox{\tiny{eff.}}}$ and
the aforementioned photon mass $M_{\gamma}$ are almost equal. Here
$\alpha$ is fixed to be $\alpha=1/137$. A qualitative analysis of
the potential (\ref{C26}) emphasizing the role played by the angle
$\theta$ will be presented in Sec. V.
\begin{figure}[hbt]
\includegraphics{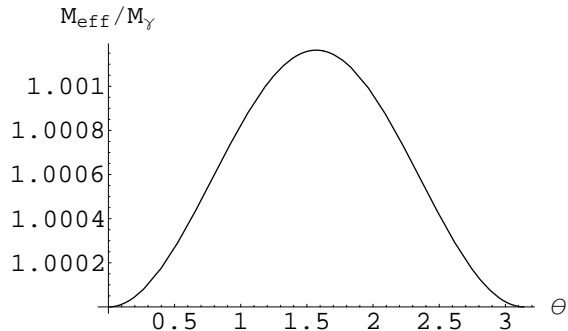}
\caption{The ratio
$\frac{M_{\mbox{\tiny{eff.}}}}{M_{\gamma}}=g(\theta)$ for $\theta\in
[0,\pi]$ in the regime $m_{dyn.}^{2}\ll
|\mathbf{q}_{\|}^{2}|\ll|eB|$ of LLL dominance. }
\end{figure}
\par\noindent
In Sec. IV we will use the Born approximation to determine the
static Coulomb potential in two regimes of dynamical mass in the LLL
approximation. Eventually we will study the behavior of the
potentials (\ref{C19}) and (\ref{C24}) as a function of the angle
$\theta$ for two limits of large and small distances.
\section{Born approximation and the static Coulomb  potential in a strong magnetic
field}\setcounter{equation}{0}\par\noindent In the nonrelativistic
quantum mechanics the relation between the scattering amplitude
${\cal{M}}$ and the potential is given by the Born approximation
\begin{eqnarray}\label{D1}
\langle p'|i{\cal{M}}|p\rangle=-i
{V}(\mathbf{q})(2\pi)\delta(E_{p'}-E_{p}),
\end{eqnarray}
where $p$ ($E_{p}$) and $p'$ ($E_{p'}$) are the momenta (energy) of
the incoming and outgoing particles, respectively, and
$\mathbf{q}=\mathbf{p}'-\mathbf{p}$. For ordinary QED with no
background magnetic field, for instance, the amplitude of a
particle-antiparticle scattering is given by \cite{peskin}
\begin{eqnarray}\label{D2}
i{\cal{M}}\sim-\frac{ie^{2}}{|\mathbf{p}-\mathbf{p'}|^{2}}.
\end{eqnarray}
Comparing with (\ref{D1}), the attractive (classical) Coulomb
potential $V(\mathbf{q})$ is thus given by
\begin{eqnarray}
V(\mathbf{q})=-\frac{e^{2}}{|\mathbf{q}|^{2}},\qquad\mbox{with}\qquad
|\mathbf{q}|\equiv |\mathbf{p}-\mathbf{p'}|.
\end{eqnarray}
After a Fourier transformation into the coordinate space, the same
potential reads
\begin{eqnarray}\label{D3}
V(\mathbf{x})=\int\frac{d^{3}q}{(2\pi)^{3}}
{V}(\mathbf{q})e^{i\mathbf{q}\cdot\mathbf{x}}=-\frac{\alpha}{R},
\end{eqnarray}
where $R\equiv |\mathbf{x}|$. Furthermore, to include the quantum
correction into the result,  the modified Coulomb potential can be
calculated from
\begin{eqnarray}\label{D4}
V(\mathbf{x})=-\int\frac{d^{3}q}{(2\pi)^{3}}
e^{i\mathbf{q}\cdot\mathbf{x}}\frac{e^{2}}{\mathbf{q}^{2}\left(1-\Pi(\mathbf{q}^{2})\right)},
\end{eqnarray}
where $\Pi(\mathbf{q})$ in the ordinary QED is defined by the vacuum
polarization tensor
\begin{eqnarray}
\Pi_{\mu\nu}(q)=\left(q^{2}g^{\mu\nu}-q^{\mu}q^{\nu}\right)\Pi({q}^{2}),
\end{eqnarray}
and is given by
\begin{eqnarray}
\Pi({q}^{2})=-\frac{2\alpha}{\pi}\int_{0}^{1}dx\
x(1-x)\log\left(\frac{m^{2}}{m^{2}-x(1-x)q^{2}}\right).
\end{eqnarray}
Choosing $q_{0}=0$ and plugging this relation into (\ref{D4}), after
some straightforward calculation \cite{peskin}, one arrives at the
so called {\it Uehling potential}
\begin{eqnarray}\label{D8}
V(R)=-\frac{\alpha}{r}\left(1+\frac{\alpha}{4\sqrt{\pi}}\frac{e^{-2mR}}{(mR)^{3/2}}+\cdots\right).
\end{eqnarray}
In this section, we will use the above method to determine the
potential of QED in a strong magnetic field in two regimes of
dynamical mass $|{\mathbf{q}}_{\|}^{2}|\ll m_{dyn.}^{2}\ll |eB|$ and
$m_{dyn.}^{2}\ll |\mathbf{q}_{\|}^{2}|\ll|eB|$ separately. We will
show that our results coincide with (\ref{C19})-(\ref{C20}) in the
regime $|{\mathbf{q}}_{\|}^{2}|\ll m_{dyn.}^{2}\ll |eB|$ and with
(\ref{C24}) in the regime $m_{dyn.}^{2}\ll
|\mathbf{q}_{\|}^{2}|\ll|eB|$.
\subsection{Modified Coulomb potential in $ |{\mathbf{q}}_{\|}^{2}|\ll m_{dyn.}^{2}\ll |eB|$ regime}
\par\noindent
We will start using the relation (\ref{D4}) where $\Pi(\mathbf{q})$
in this regime is given by (\ref{B11}). The modified Coulomb
potential in this regime is therefore given by
\begin{eqnarray}\label{D9}
V(\mathbf{x})=-e^{2}\int\frac{d^{3}q}{(2\pi)^{3}}\
\frac{e^{-i\mathbf{q}\cdot \mathbf{x}}}{\mathbf{q}^2+\frac{\gamma
eB}{2}\ {q}_{3}^{2}\ e^{-\frac{\mathbf{q}_{\perp}^{2}}{2|eB|}}},
\end{eqnarray}
where we set $q_{0}=0$, and the factor $\gamma\equiv
\frac{\alpha|eB|}{3\pi m_{dyn.}^{2}}$ is defined in (\ref{C13-a}).
The $3$-dimensional integral over $q$ can be performed by making use
of the same methods as was shown in Appendix A. The main steps of
the calculations are as follows. First using the relation
(\ref{AA2}) and (\ref{AA7}), we write the integrand in the form
\begin{eqnarray}\label{D10}
V(\mathbf{x})=-e^{2}\int\limits_{0}^{\infty}ds\int
\frac{dq_{\perp}q_{\perp}dq_{3}d\varphi'}{(2\pi)^{3}}
e^{-i\left(q_{3}R\cos\theta+q_{\perp}R\sin\theta\cos(\varphi-\varphi')\right)}\exp\left({-s\left(\mathbf{q}_{\perp}^{2}+q_{3}^{2}+\frac{\gamma
eB}{2}q_{3}^{2}e^{-\frac{\mathbf{q}_{\perp}^{2}}{2|eB|}}\right)}\right),\nonumber\\
\end{eqnarray}
where $\theta$ is the angle between the particle-antiparticle axis
and the external magnetic field. The integration over $\varphi'$ an
$q_{3}$ can be performed using (\ref{AA9}) and (\ref{AA11}),
respectively. We arrive therefore at
\begin{eqnarray}\label{D11}
V\left(\mathbf{x}\right)=-e^{2}\int\limits_{0}^{\infty}ds\sqrt{\frac{\pi}{s'}}\int\frac{dq_{\perp}q_{\perp}}{(2\pi)^{2}}J_{0}\left(q_{\perp}R\sin\theta\right)
e^{-s\mathbf{q}_{\perp}^{2}} e^{-\frac{R^{2}\cos^{2}\theta}{4s'}},
\end{eqnarray}
where we have introduced
\begin{eqnarray}\label{D12}
s'\equiv s\left(1+\frac{\gamma
eB}{2}e^{-\frac{\mathbf{q}_{\perp}^{2}}{2|eB|}}\right),
\end{eqnarray}
to simplify our notations. At this stage, to perform the integration
over $q_{\perp}$ and eventually over $s$, one can use the
approximation $|\mathbf{q}_{\perp}^{2}|\ll |eB|$, which is valid in
the LLL approximation. The factors $\frac{1}{s'}$ and
$\frac{1}{\sqrt{s'}}$ are therefore given by
\begin{eqnarray}\label{D13}
\frac{1}{s'}&\simeq&\frac{2}{s\xi}\left(1+\frac{\gamma}{2\xi}\mathbf{q}_{\perp}^{2}+\left(\frac{\gamma^{2}}{4\xi^{2}}-\frac{\gamma}{8|eB|\xi}\right)
\mathbf{q}_{\perp}^{4}\right), \nonumber\\
\frac{1}{\sqrt{s'}}&\simeq&
\sqrt{\frac{2}{s\xi}}\left(1+\frac{\gamma}{4\xi}\mathbf{q}_{\perp}^{2}+\left(\frac{3\gamma^{2}}{32\xi^{2}}-
\frac{\gamma}{16|eB|\xi}\right)\mathbf{q}_{\perp}^{4}\right),
\end{eqnarray}
where $\xi\equiv (2+\gamma|eB|)$. Using these expressions, the
potential is given by
\begin{eqnarray}\label{D14}
\lefteqn{\hspace{-1.5cm}
V(\mathbf{x})=-e^{2}\int\limits_{0}^{\infty}ds\sqrt{\frac{2\pi}{s\xi}}\int\frac{dq_{\perp}q_{\perp}}{(2\pi)^{2}}\
e^{-s\mathbf{q}_{\perp}^{2}} e^{-\frac{b_{1}}{2\xi
s}}J_{0}\left(q_{\perp}R\sin\theta\right)
}\nonumber\\
&&\hspace{-1cm}\times
\left\{1+\frac{\gamma\mathbf{q}_{\perp}^{2}}{4\xi}\left(1-\frac{b_{1}}{s\xi}
\right)
+\mathbf{q}_{\perp}^{4}\bigg[\frac{\gamma^{2}}{16\xi^{2}}\left(\frac{3}{2}-\frac{3b_{1}}{s\xi}+\frac{b_{1}^2}{2s^2\xi^2}\right)-
\frac{\gamma}{16\xi|eB|}\left(1-\frac{b_{1}}{s\xi}\right)
\bigg]\right\},
\end{eqnarray}
where $b_{1}\equiv R^{2}\cos^{2}\theta$. The result is similar to
(\ref{AA16}) and indeed we have used the same approximation which
was done in (\ref{AA15}) and led to (\ref{AA16}). Using now the
integrals $I_{i},\ i=1,2,3$ from (\ref{AA20})-(\ref{AA21-a}),
(\ref{D14}) can be written as
\begin{eqnarray}\label{D15}
\lefteqn{V(\mathbf{x})=-e^{2}\int\limits_{0}^{\infty}\frac{ds}{(2\pi)^{2}}\
\sqrt{\frac{2\pi}{s\xi}}\ e^{-\frac{b_{1}}{2\xi s}} }\nonumber\\
&&\times
\left\{I_{1}+\frac{\gamma\mathbf{q}_{\perp}^{2}}{4\xi}\left(1-\frac{b_{1}}{s\xi}
\right)I_{2}
+\mathbf{q}_{\perp}^{4}\bigg[\frac{\gamma^{2}}{16\xi^{2}}\left(\frac{3}{2}-\frac{3b_{1}}{s\xi}+\frac{b_{1}^2}{2s^2\xi^2}\right)-
\frac{\gamma}{16\xi|eB|}\left(1-\frac{b_{1}}{s\xi}\right)
\bigg]I_{3}\right\}.
\end{eqnarray}
Here, the integration over $s$ can be performed using (\ref{AA23}).
After replacing
\begin{eqnarray}\label{D16}
R^{2}\sin^{2}\theta+\frac{2R^{2}\cos^{2}\theta}{\xi}\to 4\beta
R^{2}f^{2}(\theta),
\end{eqnarray}
where $\beta\equiv (2\xi)^{-1}$ and $f^{2}(\alpha,\theta)$ are
defined in (\ref{C13-a}) and (\ref{C14}), we arrive at the same
potential (\ref{C19})-(\ref{C20}). The potential has therefore the
general form
\begin{eqnarray*}
V(R,\theta)=-\frac{\alpha}{
R}\left({\cal{A}}_{1}(\alpha,\theta)-\frac{\gamma{\cal{A}}_{2}(\alpha,\theta)}{R^{2}}+
\frac{\gamma^{2}{\cal{A}}_{3}(\alpha,\theta)}{R^{4}}\right),
\end{eqnarray*}
with ${\cal{A}}_{i}, i=1,2,3$ from (\ref{C20}).
\subsection{Modified Coulomb potential in $m_{dyn.}^{2}\ll |\mathbf{q}_{\|}^{2}|\ll|eB|$ regime}
\par\noindent
Starting from (\ref{D4}) and plugging $\Pi(q)$ corresponding to the
relevant regime $m_{dyn.}^{2}\ll |\mathbf{q}_{\|}^{2}|\ll|eB|$ from
(\ref{B12}) and choosing $q_{0}=0$, we arrive first at
\begin{eqnarray}\label{D17}
V(\mathbf{x})=-e^{2}\int\frac{d^{3}q}{(2\pi)^{3}}\
\frac{e^{-i\mathbf{q}\cdot \mathbf{x}}}{\mathbf{q}^2+\zeta^2
e^{-\frac{\mathbf{q}_{\perp}^{2}}{2|eB|}}},
\end{eqnarray}
where $\zeta\equiv \sqrt{\frac{2\alpha |eB|}{\pi}}$ is already
defined in (\ref{C23}). To perform the three dimensional integral
over $\mathbf{q}$, we will use the same methods described in
Appendix B, to evaluate the full photon propagator in the coordinate
space. First using Schwinger's parametrization technique
(\ref{AB2}), the potential reads
\begin{eqnarray}\label{D18}
\hspace{-0.5cm}V(\mathbf{x})=-e^2\int\limits_{0}^{\infty}ds\int\frac{dq_{\perp}q_{\perp}dq_{3}d\varphi'
}{(2\pi)^3}\
e^{-i(q_{\perp}R\sin\theta\cos\left(\varphi-\varphi'\right)+q_{3}R\cos\theta)}\
\exp\left(-s\left(\mathbf{q}^2+\zeta^2\
e^{-\frac{\mathbf{q}_{\perp}^{2}}{2|eB|}}\right)\right),
\end{eqnarray}
where we have used (\ref{AA7}) to bring $\mathbf{q}\cdot\mathbf{x}$
in a useful form. Now integrating over $\varphi'$ and $q_{3}$ using
(\ref{AA9}) and (\ref{AA11}), we arrive at
\begin{eqnarray}\label{D19}
V(\mathbf{x})=-e^2\int\limits_{0}^{\infty}ds
\sqrt{\frac{\pi}{s}}\int\frac{dq_{\perp}q_{\perp}}{(2\pi)^2}\
e^{-\frac{R^2\cos^2\theta}{4s}}J_{0}\left(q_{\perp}R\sin\theta\right)e^{-s\zeta^2}e^{-s\left(1-\frac{\alpha}{\pi}\right)\mathbf{q}_{\perp}^{2}},
\end{eqnarray}
where we have used the IR approximation $|\mathbf{q}|^{2}\ll |eB|$
relevant in the LLL regime
$$
e^{-s\left(\mathbf{q}_{\perp}^{2}+\zeta^{2}\exp(-\frac{\mathbf{q}^{2}_{\perp}}{2|eB|})\right)}\simeq
e^{-s\zeta^{2}}e^{-s(1-\frac{\alpha}{\pi})\mathbf{q}_{\perp}^{2}}.
$$
Next, using (\ref{AB13}) the integration over $q_{\perp}$ can be
performed. The potential is therefore given by
\begin{eqnarray*}
V(\mathbf{x})=-\frac{e^2\sqrt{\pi}}{8\pi^2\left(1-\frac{\alpha}{\pi}\right)}\int\limits_{0}^{\infty}\frac{ds}{s^{\frac{3}{2}}}\
e^{-\frac{R^2 g^{2}(\theta)}{4s}-s\zeta^2}, \qquad\mbox{with}\qquad
g^2(\theta)=\cos^{2}\theta+\frac{\sin^2\theta}{(1-\frac{\alpha}{\pi})}.
\end{eqnarray*}
The $\theta$-dependent function $g(\theta)$ is defined already in
(\ref{C23}). Finally defining a new variable $s'=\zeta^{2}s$, and
using (\ref{AB15-a}) the potential $V(\mathbf{x})$ in the
$m_{dyn.}^{2}\ll |\mathbf{q}_{\|}^{2}|\ll|eB|$ in the LLL dominance
can be calculated and reads
\begin{eqnarray*}
V(R,\theta)=-\frac{\alpha}{(1-\frac{\alpha}{\pi})g(\theta)R}e^{-\zeta
g(\theta)R},
\end{eqnarray*}
which is the same potential as (\ref{C26}) which was found using the
Wilson-loop technique.
\section{A Qualitative analysis of modified Coulomb potentials in the LLL}
\par\noindent
\setcounter{equation}{0} In this section we will consider the
potential (\ref{C20}) and (\ref{C26})
\begin{eqnarray*}
V_{1}(R,\theta)=-\frac{\alpha}{
R}\left({\cal{A}}_{1}(\alpha,\theta)-\frac{\gamma{\cal{A}}_{2}(\alpha,\theta)}{R^{2}}+
\frac{\gamma^{2}{\cal{A}}_{3}(\alpha,\theta)}{R^{4}}\right),
\end{eqnarray*}
\begin{eqnarray*}
V_{2}(R,\theta)=-\frac{\alpha}{(1-\frac{\alpha}{\pi})g(\theta)R}e^{-\zeta
g(\theta)R},
\end{eqnarray*}
in the first  $|{\mathbf{q}}_{\|}^{2}|\ll m_{dyn.}^{2}\ll |eB|$, and
the second $m_{dyn.}^{2}\ll |\mathbf{q}_{\|}^{2}|\ll|eB|$ LLL
regimes, respectively. We will eventually compare these potentials
with the modified Coulomb potential (\ref{shabad}). \par\noindent As
for $V_{1}(R,\theta)$, Figure 3 shows this potential for different
choices of the magnetic field $B=10^5, 10^6, 10^7, 10^9$ (from right
to left) and different $\theta=0,\pi/3,2\pi/3,\pi$.
\begin{center}
\par\vspace{-0.2cm}
\begin{figure}[h]
\includegraphics[width=16.5cm, height=9cm]{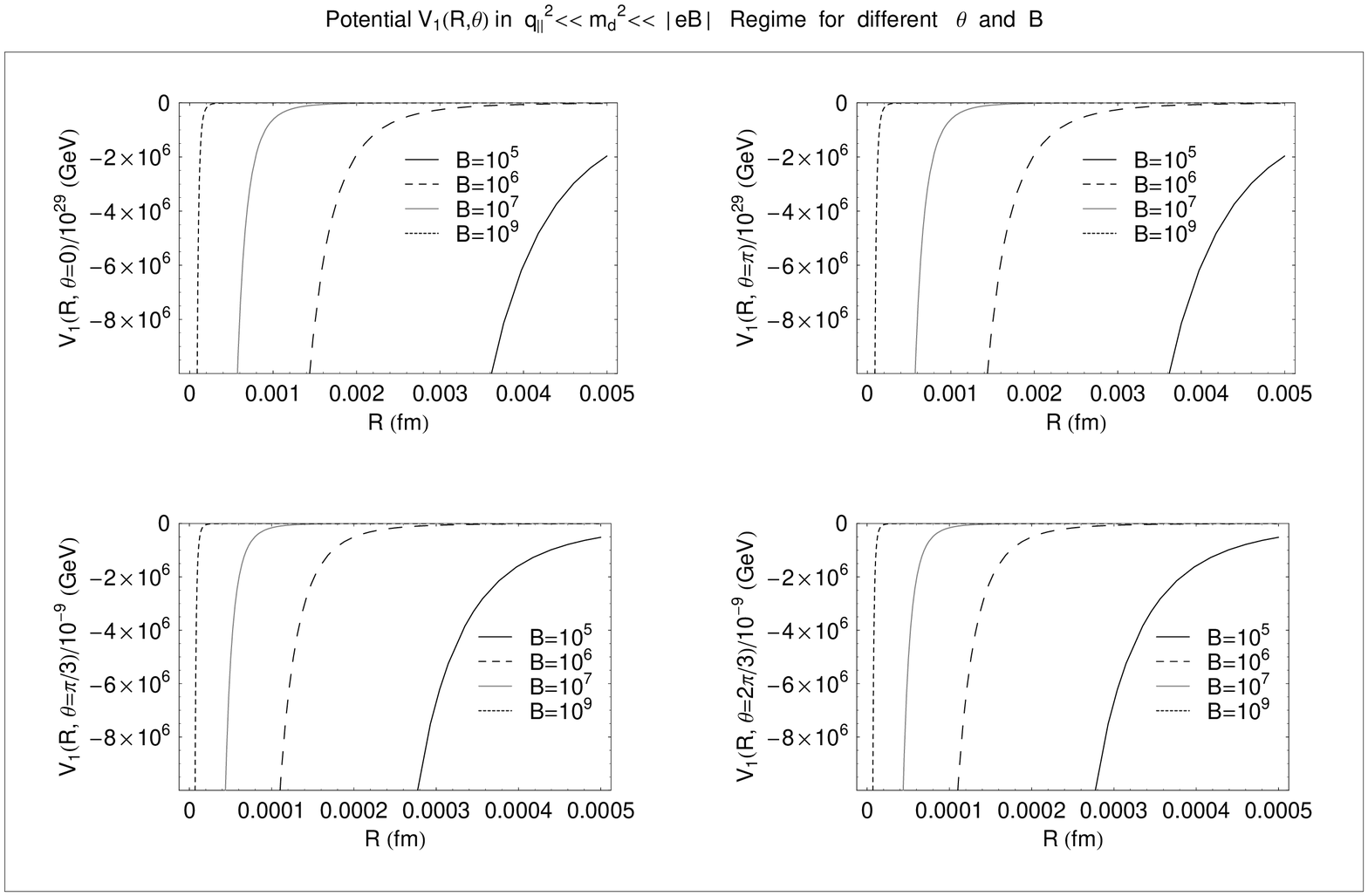}
\caption{Potential $V_{1}(R,\theta)$ for different $B$ and $\theta$.
No qualitative changes occurs by varying the angle $\theta$.}
\end{figure}
\end{center}
According to this result, for $R\to 0$ the potential falls more
rapidly to $-\infty$ the smaller the magnetic field is. Although the
scales in which the potential $V_{1}$ is plotted for $\theta=0,\pi$
are different from the scales in which it is plotted for
$\theta=\pi/3, 2\pi/3$, but their shapes are almost the same, {\it
i.e.} no qualitative changes occurs by varying the angle $\theta$.
This situation changes by neglecting the coefficient
$\gamma^{2}{\cal{A}}_{3}$ comparing with ${\cal{A}}_{1}$ and
$\gamma{\cal{A}}_{2}$ in $V_{1}(R,\theta)$. Figure 4 shows the
behavior of the coefficients ${\cal{A}}_{1}$, $\gamma{\cal{A}}_{2}$
and $\gamma^2{\cal{A}}_{3}$ of the potential $V_{1}$ as functions of
the angle $\theta$ for different magnetic fields $B=10^2, 10^5,
10^7$ and $B=10^9$. Choosing $\alpha=1/137$, it turns out that the
coefficients ${\cal{A}}_{1}, \gamma{\cal{A}}_{2}$ and
$\gamma^{2}{\cal{A}}_{3}$ are positive $\forall \theta\in [0,\pi]$
and for any choice of constant magnetic field $B$. However, as it is
shown in Figure 3, ${\cal{A}}_{3}$ decreases rapidly with increasing
magnetic field. Whereas for $B=10^2$ two coefficients
${\cal{A}}_{1},\gamma{\cal{A}}_{2}$ and $\gamma^{2}{\cal{A}}_{3}$
are comparable, for $B=10^5$, $\gamma^2{\cal{A}}_{3}$ is $10^4$
times smaller than ${\cal{A}}_{1}$, and for $B=10^9$ this difference
is $\sim 13$ order of magnitude. We conclude therefore that for
strong magnetic field $B\geq 10^5$, the coefficient
$\gamma^2{\cal{A}}_{3}$ in $V_{1}(R,\theta)$ is negligible comparing
with ${\cal{A}}_{1}$ and $\gamma{\cal{A}}_{2}$. Thus for $B\geq
10^5$ the potential $V_{1}(R,\theta)$ can be replaced by
\begin{eqnarray}\label{E-x}
V_{3}(R,\theta)=-\frac{\alpha}{
R}\left({\cal{A}}_{1}(\alpha,\theta)-\frac{\gamma{\cal{A}}_{2}(\alpha,\theta)}{R^{2}}\right),
\end{eqnarray}
which has its minimum at
\begin{eqnarray}\label{E2}
R_{min}(B,\theta)=\sqrt{\frac{3\gamma
{\cal{A}}_{2}}{{\cal{A}}_{1}}}.
\end{eqnarray}
\begin{center}
\begin{figure}[t]
\includegraphics[width=14cm,height=3cm]{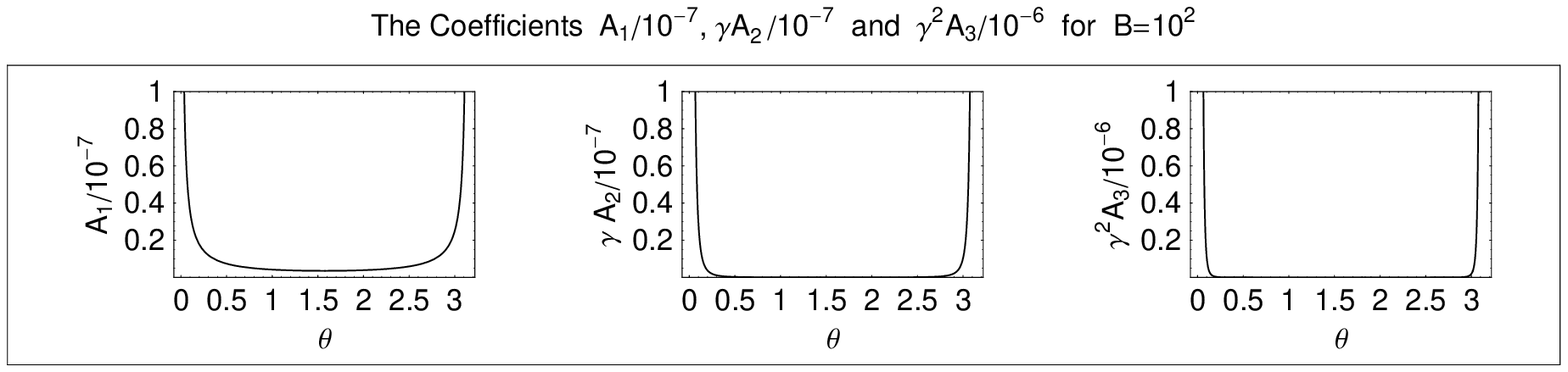}
\includegraphics[width=14cm,height=3cm]{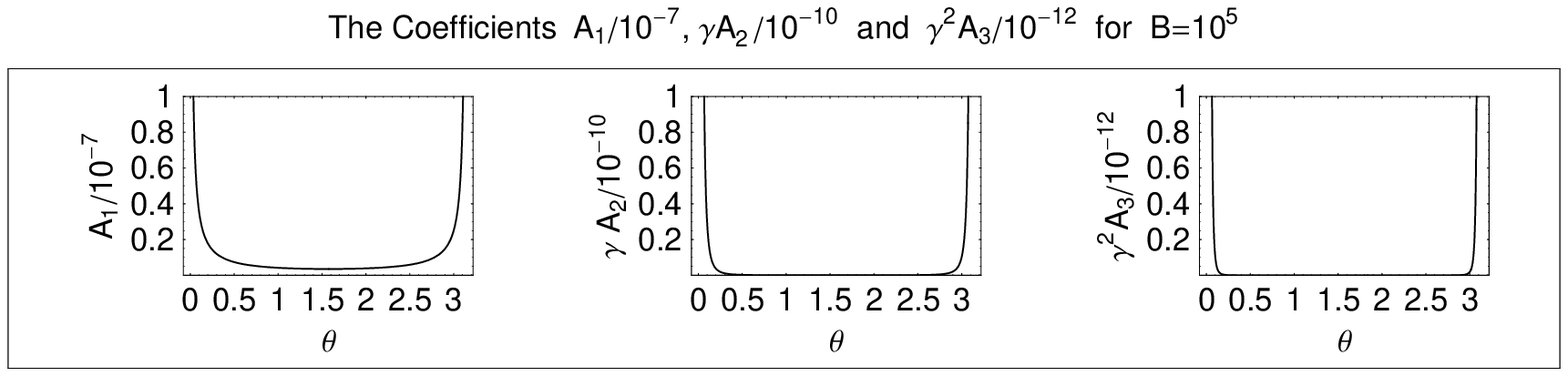}
\includegraphics[width=14cm,height=3cm]{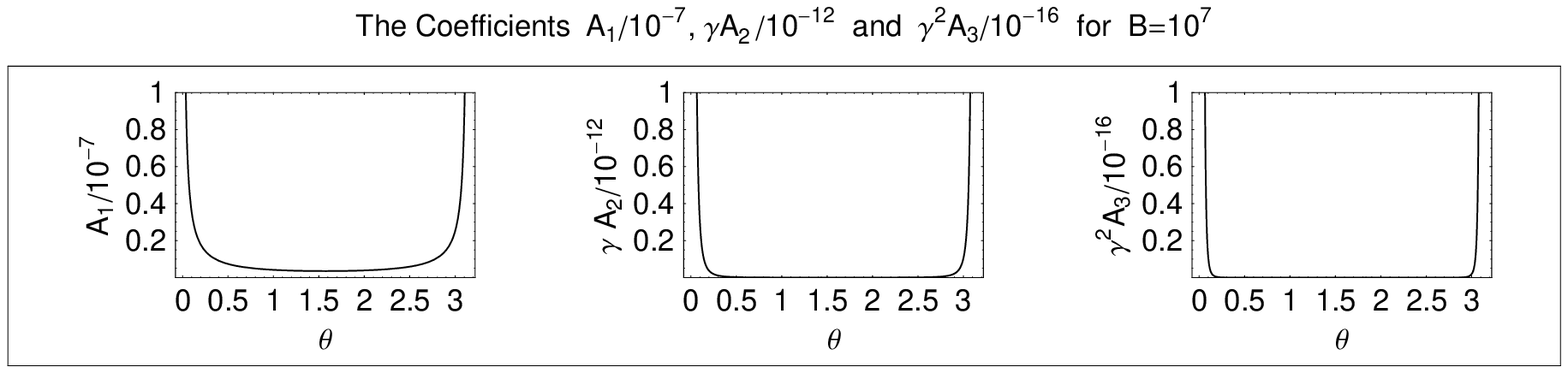}
\includegraphics[width=14cm,height=3cm]{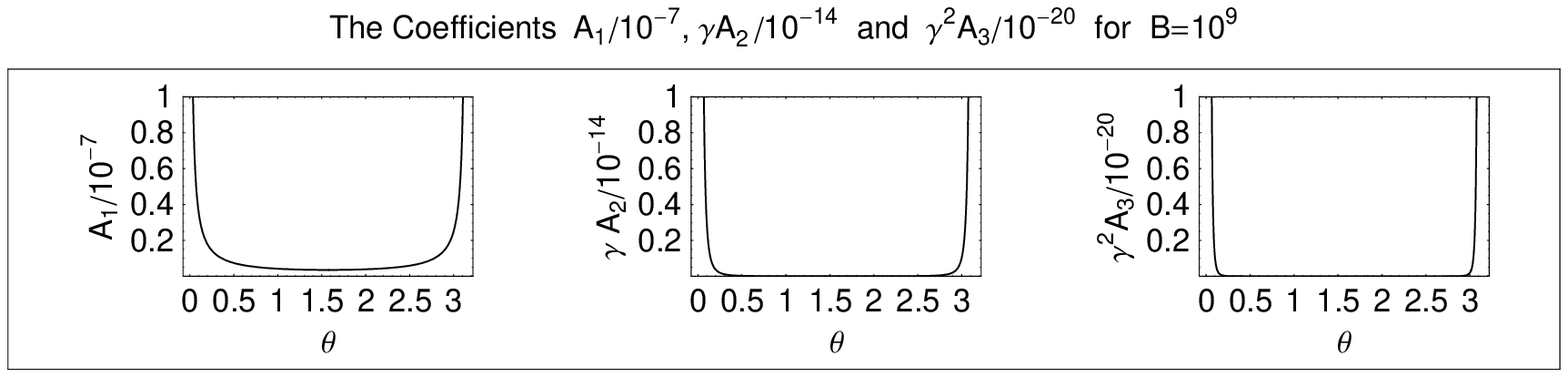}
\caption{Coefficients ${\cal{A}}_{1}$, $\gamma{\cal{A}}_{2}$ and
$\gamma^2{\cal{A}}_{3}$ for different magnetic fields. As it turns
out, ${\cal{A}}_{3}$ decreases rapidly with increasing magnetic
field.}
\end{figure}
\end{center}
Figure 5 shows $V_{3}(R,\theta)$ for different choices of the
magnetic field $B=10^5, 10^6, 10^7$ and $10^9$ (from right to left)
and different $\theta=0, \pi/3, \pi/2, 2\pi/3,\pi$. Whereas for
$\theta=0,\pi$ the potential is repulsive, it exhibits a minimum for
angles $\theta\in]0,\pi[$ and distances $R\leq 0.005$ fm. The depth
of the potential at $R_{min}$ increases with the magnetic field. We
interpret this effect as a possibility for bound state formation.
The answer to the question concerning the existence and the number
of bound states in the potential $V_{3}$ is beyond the scope of this
paper.
\begin{center}
\begin{figure}[h]
\includegraphics[width=16.5cm, height=9.5cm]{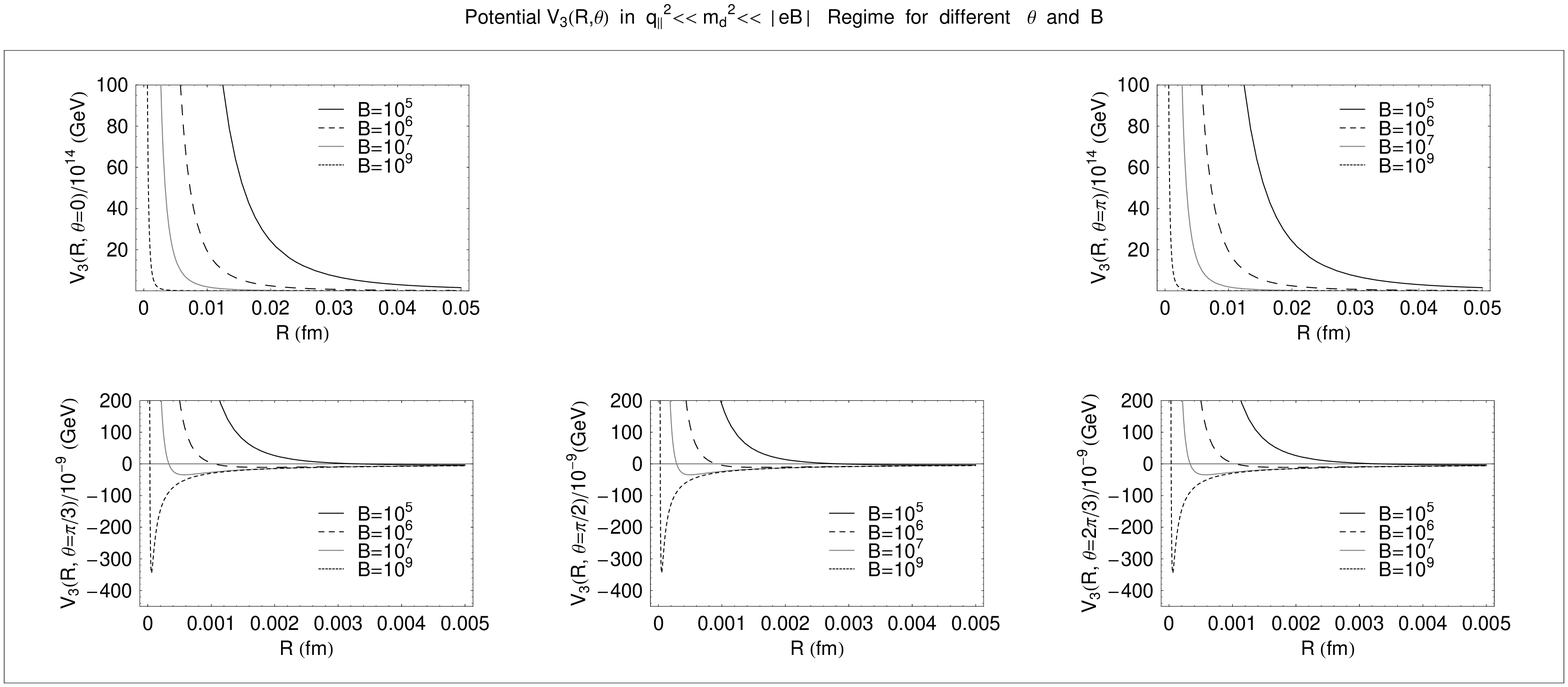}
\caption{Potential $V_{3}(R,\theta)$ for different $B$ and $\theta$.
Bound states can be formed for $\theta\in ]0,\pi[$ and for strong
magnetic fields $B\geq 10^5$ in the regime $R\leq 0.005$ fm. The
depth of the potential at $R_{min}$ increases with the magnetic
field.}
\end{figure}
\end{center}
In Figure 6 the behavior of $R_{min}$ from (\ref{E2}) for different
$\theta$ is studied. As it turns out, for different $\theta$, the
position of the minimum of the potential is proportional to
$1/\sqrt{B}$.
\begin{center}
\begin{figure}[h]
\includegraphics[width=16.5cm, height=6cm]{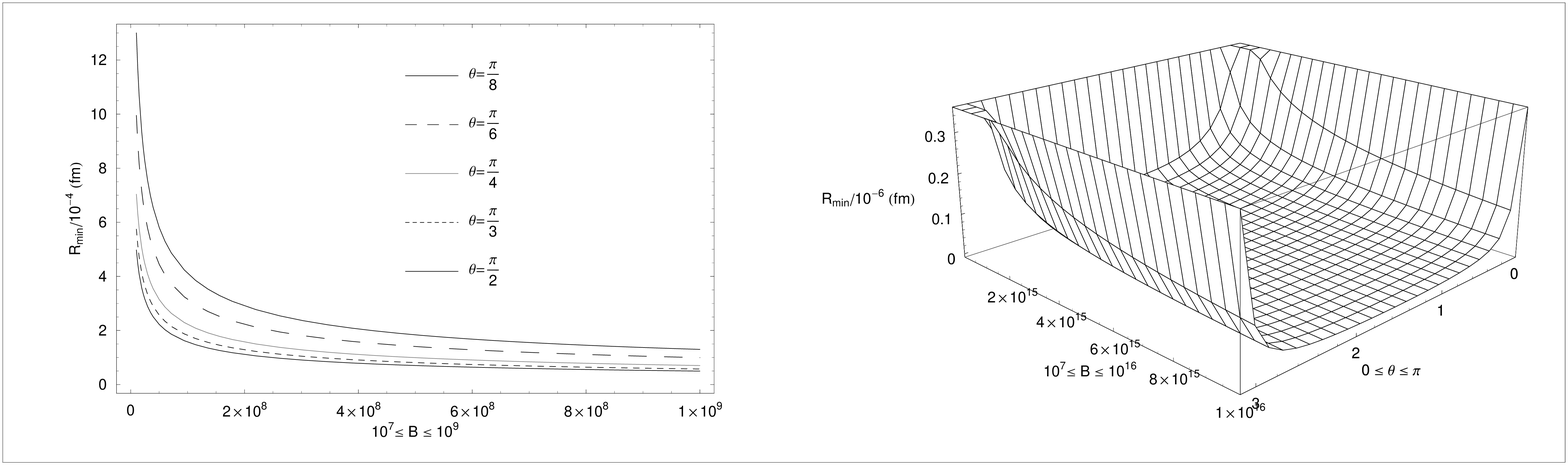}
\caption{Behavior of the minimum of the potential $V_{3}$ as a
function of $10^7\leq B\leq 10^{16}$ for different $0<\theta\leq
\pi/2$. For $\pi/2\leq \theta<\pi$, $R_{min}$ shows a symmetry in
changing $\theta\to\theta+\pi/2$ (see the 3-dimensional figure on
the r.h.s.).}
\end{figure}
\end{center}
\newpage
\begin{figure}[t]
\includegraphics[width=16.5cm, height=11cm]{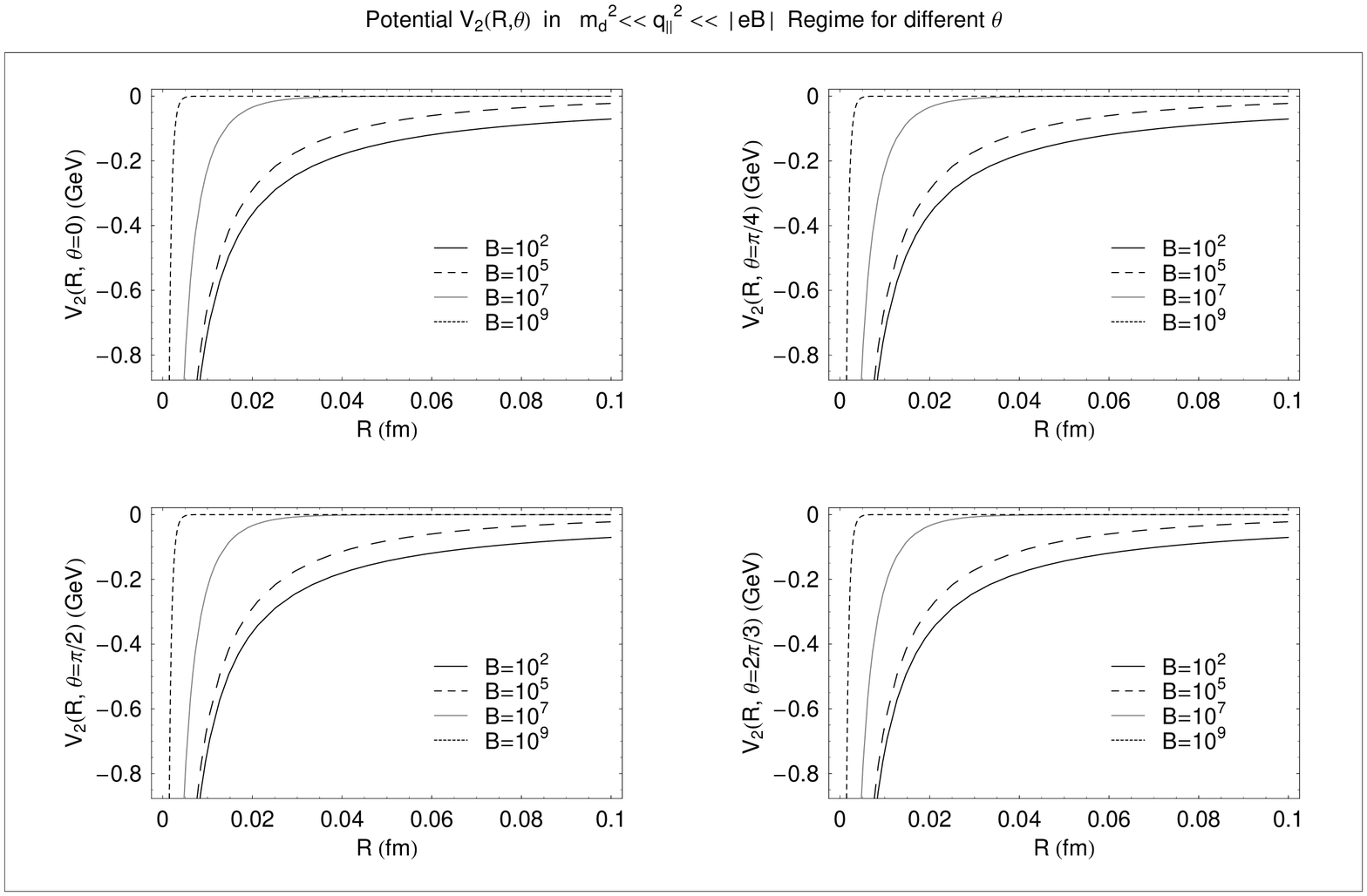}
\caption{Potential $V_{2}(R,\theta)$ for different $B$ and $\theta$.
No qualitative changes occurs by varying the angle $\theta$. This
potential is comparable with the potential (\ref{shabad}) from
\cite{shabad-1} in the scaling regime.}
\end{figure}
Let us now consider the potential $V_{2}(R,\theta)$ in the second
regime $m_{dyn.}^{2}\ll |\mathbf{q}_{\|}^{2}|\ll|eB|$ of LLL. Figure
7 shows its behavior for different magnetic field $B$ and angle
$\theta$. Again no qualitative changes occurs by varying the angle
$\theta$. As it is pointed out in the introduction, this Yukawa-like
potential is comparable with the potential (\ref{shabad}) from
\cite{shabad-1} in the scaling regime.
\section{Summary}
\par\noindent
In this paper the static potential of QED is calculated in the
presence of a strong but constant magnetic field using two different
methods. First a perturbative Wilson loop calculation is performed
for two different regimes of dynamical mass
$|{\mathbf{q}}_{\|}^{2}|\ll m_{dyn.}^{2}\ll |eB|$, and
$m_{dyn.}^{2}\ll |\mathbf{q}_{\|}^{2}|\ll|eB|$ in the lowest Landau
level (LLL). The resulting potential is then compared with the
potential arising from a modified Born approximation. The results
coincide. Comparing with the recently calculated potential of point
like charges in \cite{shabad-1}, our potential shows a novel
dependence on the angle $\theta$ between the particle-antiparticle
axis and the direction of the magnetic field. A qualitative analysis
of these modified potentials is performed in the previous Sec. V. As
for the potential (\ref{C20}) from the first regime
$|{\mathbf{q}}_{\|}^{2}|\ll m_{dyn.}^{2}\ll |eB|$, it turns out that
for strong magnetic field $B\geq 10^5$, the coefficient of the
$R^{-5}$ term, {\it i.e.} $\gamma^2 {\cal{A}}_{3}$, is negligible
comparing to the both other coefficients ${\cal{A}}_{1}$ and $\gamma
{\cal{A}}_{2}$ from the expected Coulombian $R^{-1}$ and the
additional $R^{-3}$ terms. Keeping only these two coefficients, a
qualitative change occurs in the Coulomb-like potential which
depends on the angle $\theta$; Whereas for $\theta=0,\pi$ the
potential is repulsive, it exhibits a minimum for angles
$\theta\in]0,\pi[$ and distances $R\leq 0.005$ fm. The depth of the
potential at $R_{min}$ increases with the magnetic field and the
position of the minimum is proportional to $1/\sqrt{B}$. We
interpret this effect as a possibility for bound state formation. A
rigorous proof of the existence and the number of bound states in
the above potential is the subject of future investigations.
\par
As for the potential (\ref{C26}) of the second regime
$m_{dyn.}^{2}\ll |\mathbf{q}_{\|}^{2}|\ll|eB|$, it is comparable
with a screened Yukawa potential. It is well-known that in this
regime the photon acquires a finite mass proportional to
$\sqrt{|eB|}$. According to our result, the photon receives also a
modified effective mass depending on the angle $\theta$ between the
particle axis and the direction of the magnetic field. But, this
dependence is indeed negligible for fixed $\alpha=1/137$. This
result is in good agreement with the result recently found in
\cite{shabad-1}.

It would be interesting to study the renormalization group (RG)
improvement of these potentials, as they both depend on the coupling
constant $e$, which is taken to be a bare parameter in our
calculations. Using the RG improved potential it is possible,
according to \cite{kogut-paper}, to determine the Callan-Symanzik
$\beta$-function of QED in the presence of strong magnetic field in
the LLL approximation.
\section*{Acknowledgments}
The authors thank F. Ardalan and H. Arfaei for useful discussions.
N.S. thanks in particular the referee of the paper for valuable
hints concerning the potential $V(x)$ from \cite{miransky1-5}.
\begin{appendix}
\section{The LLL photon propagator in $ |{\mathbf{q}}_{\|}^{2}|\ll m_{dyn.}^{2}\ll |eB|$ regime}
\par\noindent
In this section we will perform the integration over $q$ in
(\ref{C12-a}) to determine the full LLL photon propagator in
$m_{dyn.}^{2}\ll |\mathbf{q}_{\|}^{2}|\ll|eB|$ regime (\ref{C13}).
To start, let us give the integral (\ref{C12-a}) in the Euclidean
space
\begin{eqnarray}\label{AA1}
\widetilde{\cal{D}}_{\mu\nu}(x)=\delta_{\mu\nu}^{\|}\int\frac{d^{4}q}{(2\pi)^{4}}
\frac{e^{-i(q_{4}x_{4}+\mathbf{q}\cdot\mathbf{x})}}{q^{2}+
\frac{\gamma(\alpha)|eB|}{2}\left({q}_{3}^{2}+
q_{4}^{2}\right)\exp\left(-\frac{\mathbf{q}_{\perp}^{2}}{2|eB|}\right)},
\end{eqnarray}
where $\gamma(\alpha)\equiv \frac{2\alpha}{3\pi m_{dyn.}^{2}}$.
Here, the Euclidean coordinates $x_{0}=-ix_{4}$ as well as
$q_{0}=-iq_{4}$. We have further used the notations
$\mathbf{q}=(q_{1},q_{2},q_{3})$ and
$\mathbf{x}=(x_{1},x_{2},x_{3})$. The scalar product is defined
therefore by $\mathbf{q}\cdot \mathbf{x}=\sum_{i=1}^{3}q_{i}x_{i}$.
Using the Schwinger parametrization technique
\begin{eqnarray}\label{AA2}
\int_{0}^{\infty}ds\ e^{-a s}=\frac{1}{a},
\end{eqnarray}
the above expression (\ref{AA1}) can be given by
\begin{eqnarray}\label{AA3}
\widetilde{\cal{D}}_{\mu\nu}(x)=\delta_{\mu\nu}^{\|}\int_{0}^{\infty}ds\int\frac{d^{4}q}{(2\pi)^{4}}
{e^{-i(q_{4}x_{4}+\mathbf{q}\cdot\mathbf{x})}}\exp\left(
-s\left(q_{4}^{2}+\mathbf{q}^{2}+
\frac{\gamma(\alpha)|eB|}{2}\left({q}_{3}^{2}+
q_{4}^{2}\right)\exp\left(-\frac{\mathbf{q}_{\perp}^{2}}{2|eB|}\right)\right)\right).\nonumber\\
\end{eqnarray}
Let us first evaluate the integral over $q_{4}$
\begin{eqnarray}\label{AA4}
\int_{-\infty}^{+\infty}dq_{4}e^{-iq_{4}x_{4}}\exp\left({-sq_{4}^{2}\left(1+
\frac{\gamma(\alpha)|eB|}{2}e^{-\frac{\mathbf{q}^{2}_{\perp}}{2|eB|}}\right)}\right).
\end{eqnarray}
For the new variable
\begin{eqnarray}\label{AA4s}
s'\equiv s \left(1+
\frac{\gamma(\alpha)|eB|}{2}e^{-\frac{\mathbf{q}^{2}_{\perp}}{2|eB|}}\right),
\end{eqnarray}
we get
\begin{eqnarray}\label{AA5}
\int_{-\infty}^{+\infty}dq_{4}e^{-iq_{4}x_{4}-s'q_{4}^{2}}=\sqrt{\frac{\pi}{s'}}e^{-\frac{x_{4}^{2}}{4s'}}.
\end{eqnarray}
To integrate the variable ${q}_{3}$ component, let us go into the
polar coordinate system. Here, the product
$\mathbf{q}\cdot\mathbf{x}$ is given by
\begin{eqnarray}\label{AA6}
\mathbf{q}\cdot\mathbf{x}=qR\sin\theta\sin\theta'\cos\left(\varphi-\varphi'\right)+q_{3}R\cos\theta'\cos\theta,
\end{eqnarray}
with $q\equiv |\mathbf{q}|$ and $R\equiv |\mathbf{x}|$, $\theta$
($\theta'$) the angle between $\mathbf{x}$ ($\mathbf{q}$) and the
external magnetic field, which is assumed to be in the
$x_{3}$-direction. Now using $q_{3}=q\cos\theta'$ and
$\mathbf{q}_{\perp}=q\sin\theta'$, (\ref{AA5}) is given by
\begin{eqnarray}\label{AA7}
\mathbf{q}\cdot\mathbf{x}=q_{\perp}R\sin\theta\cos\left(\varphi-\varphi'\right)+q_{3}R\cos\theta.
\end{eqnarray}
In these cylindric coordinates the photon propagator (\ref{AA3}) is
given by
\begin{eqnarray}\label{AA8}
\widetilde{\cal{D}}_{\mu\nu}(x)&=&\delta_{\mu\nu}^{\|}\int_{0}^{\infty}ds\int\frac{dq_{\perp}q_{\perp}dq_{3}d\varphi'}{(2\pi)^{4}}
e^{-i\left(q_{3}R\cos\theta+q_{\perp}R\sin\theta\cos(\varphi'-\varphi)
\right)}\nonumber\\
&& \times\exp\left( -s\left(q_{3}^{2}+\mathbf{q}_{\perp}^{2}+
\frac{\gamma(\alpha)|eB|}{2}{q}_{3}^{2}\exp\left(-\frac{\mathbf{q}_{\perp}^{2}}{2|eB|}\right)\right)\right)
\sqrt{\frac{\pi}{s'}}e^{-\frac{x_{4}^{2}}{4s'}},
\end{eqnarray}
where we have inserted the expression from (\ref{AA5}). Now using
the integral representation of the Bessel function $J_{0}$
\begin{eqnarray}\label{AA9}
J_{0}(z)=\frac{1}{2\pi}\int_{0}^{2\pi}d\varphi'
e^{-iz\cos(\varphi'-\varphi)},
\end{eqnarray}
and performing the integration over $\varphi$ we get
\begin{eqnarray}\label{AA10}
\widetilde{\cal{D}}_{\mu\nu}(x)&=&\delta_{\mu\nu}^{\|}\int_{0}^{\infty}ds\int\frac{dq_{\perp}q_{\perp}dq_{3}}{(2\pi)^{3}}
e^{-iq_{3}R\cos\theta-s'q_{3}^{2}}
J_{0}\left(q_{\perp}R\sin\theta\right)e^{ -s\mathbf{q}_{\perp}^{2}}
\sqrt{\frac{\pi}{s'}}e^{-\frac{x_{4}^{2}}{4s'}}.
\end{eqnarray}
Performing now the integration over $q_{3}$ using
\begin{eqnarray}\label{AA11}
\int_{-\infty}^{+\infty}dq_{3}e^{-iq_{3}R\cos\theta-s'q_{3}^{2}}=\sqrt{\frac{\pi}{s'}}e^{-\frac{R^{2}\cos^{2}\theta}{4s'}}.
\end{eqnarray}
The photon propagator (\ref{AA10}) therefore reads
\begin{eqnarray}\label{AA12}
\widetilde{\cal{D}}_{\mu\nu}(x)&=&\delta_{\mu\nu}^{\|}\int_{0}^{\infty}\frac{ds}{s'}\int\frac{dq_{\perp}q_{\perp}}{8\pi^{2}}
J_{0}\left(q_{\perp}R\sin\theta\right)e^{ -s\mathbf{q}_{\perp}^{2}}
e^{-\frac{\left(x_{4}^{2}+R^{2}\cos^{2}\theta\right)}{4s'}}.
\end{eqnarray}
To perform the integration over $q_{\perp}$ we proceed as follows.
Using first the IR approximation $\mathbf{q}_{\perp}^{2}\ll |eB|$ in
the regime of LLL dominance, and expanding the factor
$e^{-\frac{\mathbf{q}_{\perp}^{2}}{2|eB|}}\approx
1-\frac{\mathbf{q}_{\perp}^{2}}{2|eB|}+\frac{1}{2!}\frac{\mathbf{q}_{\perp}^{4}}{4(eB)^2}$,
the relation $s'/s$ from (\ref{AA4s}) can be written as
\begin{eqnarray}\label{AA14}
\frac{1}{s'}=\frac{2}{s\xi}
\left(1+\frac{\gamma\mathbf{q}_{\perp}^{2}}{2\xi}+\left(\frac{\gamma^{2}}{4\xi^2}-\frac{\gamma}{8\xi|eB|}\right)\mathbf{q}_{\perp}^{4}\right),
\qquad\mbox{with}\qquad \xi\equiv(2+\gamma|eB|).
\end{eqnarray}
Plugging this result in (\ref{AA12}) the photon propagator is given
by
\begin{eqnarray}\label{AA15}
\hspace{-1cm}\widetilde{\cal{D}}_{\mu\nu}(x)&=&\delta_{\mu\nu}^{\|}\int_{0}^{\infty}\frac{ds}{s\xi}\int\frac{dq_{\perp}q_{\perp}}{4\pi^{2}}
J_{0}\left(q_{\perp}R\sin\theta\right)e^{
-s\mathbf{q}_{\perp}^{2}}\left(1+\frac{\gamma\mathbf{q}_{\perp}^{2}}{2\xi}+
\left(\frac{\gamma^{2}}{4\xi^2}-\frac{\gamma}{8\xi|eB|}\right)\mathbf{q}_{\perp}^{4}\right)\nonumber\\
&&\times
e^{-\frac{\left(x_{4}^{2}+R^{2}\cos^{2}\theta\right)}{2s\xi}\left(1+\frac{\gamma\mathbf{q}_{\perp}^{2}}{2\xi}+\left(\frac{\gamma^{2}}{4\xi^2}-
\frac{\gamma}{8\xi|eB|}\right)\mathbf{q}_{\perp}^{4}\right)}.
\end{eqnarray}
Again using the IR approximation and expanding the exponent we get
\begin{eqnarray}\label{AA16}
\lefteqn{\hspace{-1.cm}\widetilde{\cal{D}}_{\mu\nu}(x)=\frac{\delta_{\mu\nu}^{\|}}{4\pi^{2}\xi}\int_{0}^{\infty}\frac{ds}{s}
e^{-\frac{b_{1}}{2s\xi}}\int_{0}^{\infty} dq_{\perp}q_{\perp}
J_{0}\left(q_{\perp}R\sin\theta\right)e^{ -s\mathbf{q}_{\perp}^{2}}}\nonumber\\
&&\hspace{-1cm}\times
\left\{1+\frac{\gamma\mathbf{q}_{\perp}^{2}}{2\xi}\left(1-\frac{b_{1}}{2s\xi}
\right)
+\mathbf{q}_{\perp}^{4}\bigg[\frac{\gamma^{2}}{4\xi^{2}}\left(1-\frac{b_{1}}{s\xi}+\frac{b_{1}^2}{8s^2\xi^2}\right)-
\frac{\gamma}{8\xi|eB|}\left(1-\frac{b_{1}}{2s\xi}\right)
\bigg]\right\},
\end{eqnarray}
where
\begin{eqnarray}
b_{1}\equiv x_{4}^{2}+R^{2}\cos^{2}\theta.
\end{eqnarray}
To perform the integration over $q_{\perp}$ we use \cite{gradshteyn}
\begin{eqnarray}\label{AA17}
\int_{0}^{\infty}dz\ z^{\mu}e^{-\kappa z^{2}}J_{\nu}\left(\beta
z\right)=\frac{\Gamma\left(\frac{\mu+\nu+1}{2}\right)}{\beta\kappa^{\frac{\mu}{2}}\Gamma\left(\nu+1\right)}e^{-\frac{\beta^{2}}{8\kappa}}
M_{\frac{\mu}{2},\frac{\nu}{2}}\left(\frac{\beta^{2}}{4\kappa}\right),
\end{eqnarray}
where $M_{\frac{\mu}{2},\frac{\nu}{2}}\left(y\right)$ is the
Whittaker function defined by
\begin{eqnarray}\label{AA18}
M_{\frac{\mu}{2},\frac{\nu}{2}}\left(y\right)=y^{\frac{\nu+1}{2}}e^{-\frac{y}{2}}\Phi\left(\frac{\nu-\mu+1}{2},\nu+1;y\right),
\end{eqnarray}
where
\begin{eqnarray}\label{AA19}
\Phi(\eta,\tau;y)\equiv
1+\frac{\eta}{\tau}\frac{y}{1!}+\frac{\eta(\eta+1)}{\tau(\tau+1)}\frac{y^{2}}{2!}+\cdots.
\end{eqnarray}
For the first term on the second line of (\ref{AA16}) we get
therefore
\begin{eqnarray}\label{AA20}
I_{1}(s)\equiv\int_{0}^{\infty}dq_{\perp}q_{\perp}\
J_{0}\left(q_{\perp}
R\sin\theta\right)e^{-s\mathbf{q}^{2}_{\perp}}=\frac{1}{2s}e^{-\frac{R^{2}\sin^{2}\theta}{4s}},
\end{eqnarray}
and for the second term we get
\begin{eqnarray}\label{AA21}
I_{2}(s)\equiv\int_{0}^{\infty}dq_{\perp}q^{3}_{\perp}\
J_{0}\left(q_{\perp}
R\sin\theta\right)e^{-s\mathbf{q}^{2}_{\perp}}=\frac{1}{2s^{2}}e^{-\frac{R^{2}\sin^{2}\theta}{4s}}\left(1-\frac{R^{2}\sin^{2}\theta}{4s}\right),
\end{eqnarray}
whereas the third term yields
\begin{eqnarray}\label{AA21-a}
I_{3}(s)\equiv\int_{0}^{\infty}dq_{\perp}q^{5}_{\perp}\
J_{0}\left(q_{\perp}
R\sin\theta\right)e^{-s\mathbf{q}^{2}_{\perp}}=\frac{1}{s^{3}}e^{-\frac{R^{2}\sin^{2}\theta}{4s}}\left(1-\frac{R^{2}\sin^{2}\theta}{2s}+
\frac{R^{4}\sin^{4}\theta}{32s^{2}}\right).
\end{eqnarray}
Inserting (\ref{AA20}), (\ref{AA21}), and (\ref{AA21-a}) in
(\ref{AA16}) we arrive first at
\begin{eqnarray}\label{AA22}
\widetilde{\cal{D}}_{\mu\nu}(x)&=&\frac{\delta_{\mu\nu}^{\|}}{4\pi^{2}\xi}\int_{0}^{\infty}\frac{ds}{s}
e^{-\frac{b_{1}}{2s\xi}}
\left\{I_{1}(s)+\frac{\gamma}{2\xi}\left(1-\frac{b_{1}}{2s\xi}\right)I_{2}(s)\right.
\nonumber\\
&&\left.+\bigg[\frac{\gamma^{2}}{4\xi^{2}}\left(1-\frac{b_{1}}{s\xi}+\frac{b_{1}^2}{8s^2\xi^2}\right)-
\frac{\gamma}{8\xi|eB|}\left(1-\frac{b_{1}}{2s\xi}\right)
\bigg]I_{3}(s) \right\}.
\end{eqnarray}
Finally, the integration over $s$ can be evaluated using
\cite{gradshteyn}
\begin{eqnarray}\label{AA23}
\int_{0}^{\infty}\frac{ds}{s^{n}}e^{-\frac{a}{s}}=\frac{\Gamma(n-1)}{a^{n-1}}.
\end{eqnarray}
Choosing the notation
$b_{0}=R^{2}\sin^{2}\theta+\frac{2b_{1}}{\xi}$, we get for the first
term in (\ref{AA22})
\begin{eqnarray}\label{AA24}
\frac{\delta^{\|}_{\mu\nu}}{8\pi^{2}\xi}\int\frac{ds}{s^{2}}e^{-\frac{b_{0}}{4s}}
=\frac{\delta_{\mu\nu}^{\|}}{2\pi^{2}}\frac{1}{\xi b_{0}},
\end{eqnarray}
for the second term
\begin{eqnarray}\label{AA25}
\frac{\delta_{\mu\nu}^{\|}\gamma}{8\pi^{2}\xi^{2}}\int_{0}^{\infty}\frac{ds}{s}
e^{-\frac{b_{0}}{4s}}\left(1-\frac{b_{1}}{2s\xi}
\right)I_{2}(s)=\frac{\delta_{\mu\nu}^{\|}\gamma}{\pi^{2}}\left(-\frac{4b_{1}}{\xi^{3}b_{0}^{3}}+\frac{12b_{1}R^{2}\sin^{2}\theta}{\xi^{3}b_{0}^{4}}+\frac{1}{\xi^{2}b_{0}^{2}}
-\frac{2R^{2}\sin^{2}\theta}{\xi^{2}b_{0}^{3}}\right),
\end{eqnarray}
and the third term
\begin{eqnarray}\label{AA25-a}
\lefteqn{\hspace{-1cm}-\frac{\delta_{\mu\nu}^{\|}}{4\pi^{2}}\int_{0}^{\infty}\frac{ds}{s^{2}}e^{-\frac{b_{0}}{4s}}\bigg[\frac{\gamma^{2}}{4\xi^{2}}\left(1-\frac{b_{1}}{s\xi}+\frac{b_{1}^2}{8s^2\xi^2}\right)-
\frac{\gamma}{8\xi|eB|}\left(1-\frac{b_{1}}{2s\xi}\right) \bigg]
I_{3}(s) = }\nonumber\\
&&\hspace{-1.2cm}=
\frac{8\delta_{\mu\nu}^{\|}\gamma^{2}}{\pi^{2}\xi^3
b_{0}^3}\left(1-\frac{6R^2\sin^{2}\theta}{b_{0}}+\frac{6R^4\sin^{4}\theta}{b_{0}^{2}}\right)
\nonumber\\
&&\hspace{-1.3cm}-\frac{96\delta^{\|}_{\mu\nu}\gamma^{2}b_{1}}{\pi^{2}\xi^{4}b_{0}^{4}}\bigg[1-\frac{8}{b_{0}}\left(R^{2}\sin^{2}\theta+\frac{b_{1}}{4\xi}\right)
+\frac{20}{b_{0}^{2}}\left(\frac{R^{4}\sin^{4}\theta}{2}+\frac{R^{2}b_{1}\sin^{2}\theta}{\xi}\right)
-\frac{30b_{1}R^{4}\sin^{4}\theta}{\xi
b_{0}^{3}}\bigg]\nonumber\\
&&\hspace{-1.3cm}-\frac{4\delta_{\mu\nu}^{\|}\gamma}{\pi^{2}|eB|\xi^{2}b_{0}^3}\bigg[1-\frac{6}{b_{0}}\left(R^{2}\sin^{2}\theta+\frac{b_{1}}{\xi}\right)+
\frac{6}{b_{0}^{2}}\left(R^{4}\sin^{4}\theta+\frac{8b_{1}R^{2}\sin^{2}\theta}{\xi}\right)-\frac{60b_{1}R^{4}\sin^{4}\theta}{\xi
b_{0}^{3}}\bigg].
\end{eqnarray}
Further, to find the LLL photon propagator in the regime
$|{\mathbf{q}}_{\|}^{2}|\ll m_{dyn.}^{2}\ll |eB|$ from (\ref{C13}),
we will first replace
\begin{eqnarray}
b_{0}\to 4\beta a_{1}(R,\theta,x_{4}),\qquad b_{1}\to
\frac{a_{2}(R,\theta,x_{4})}{\beta\gamma}\qquad\mbox{and} \qquad
\xi\to(2\beta)^{-1},
\end{eqnarray}
where $\beta$ and $a_{i},i=1,2$ are defined in (\ref{C14}). Adding
then the results from (\ref{AA24}), (\ref{AA25}) and (\ref{AA25-a})
together we arrive at the propagator  (\ref{C13}).
\section{The LLL photon propagator in $m_{dyn.}^{2}\ll |\mathbf{q}_{\|}^{2}|\ll|eB|$ regime}
\par\noindent
In this section we will perform the integration over $q$ in
(\ref{C21}) to determine the full LLL photon propagator in
$m_{dyn.}^{2}\ll |\mathbf{q}_{\|}^{2}|\ll|eB|$ regime (\ref{C22}).
To start, let us give the integral (\ref{C21}) in the Euclidean
space
\begin{eqnarray}\label{AB1}
\widetilde{\cal{D}}_{\mu\nu}(x)=
\delta_{\mu\nu}^{\|}\int\frac{d^{4}q}{(2\pi)^{4}}\
\frac{e^{-i(q_{4}x_{4}+\mathbf{q}\cdot\mathbf{x})}}{q_{4}^{2}+
\mathbf{q}^{2}+\frac{2\alpha|eB|}{\pi}e^{-\frac{\mathbf{q}_{\perp}^{2}}{2|eB|}}},
\end{eqnarray}
where we have introduced the Euclidean coordinates $x_{0}=-ix_{4}$
as well as $q_{0}=-iq_{4}$. Here, $\mathbf{q}=(q_{1},q_{2},q_{3})$
and $\mathbf{x}=(x_{1},x_{2},x_{3})$ and $\mathbf{q}\cdot
\mathbf{x}=\sum_{i=1}^{3}q_{i}x_{i}$. Using the Schwinger
parametrization
\begin{eqnarray}\label{AB2}
\int_{0}^{\infty}ds\ e^{-a s}=\frac{1}{a},
\end{eqnarray}
the above expression (\ref{AB1}) can be given by
\begin{eqnarray}\label{AB3}
\widetilde{\cal{D}}_{\mu\nu}(x)=
\delta_{\mu\nu}^{\|}\int_{0}^{\infty}
ds\int\frac{d^{4}q}{(2\pi)^{4}}\
e^{-i(q_{4}x_{4}+\mathbf{q}\cdot\mathbf{x})}\
\exp\left(-s\left(q_{4}^{2}+
\mathbf{q}^{2}+\frac{2\alpha|eB|}{\pi}e^{-\frac{\mathbf{q}_{\perp}^{2}}{2|eB|}}\right)\right).
\end{eqnarray}
Performing the integration over $q_{4}$
\begin{eqnarray}\label{AB4}
\int_{-\infty}^{+\infty}dq_{4}\
e^{-sq_{4}^{2}-iq_{4}x_{4}}=\sqrt{\frac{\pi}{s}}e^{-\frac{x_{4}^{2}}{4s}},
\end{eqnarray}
we arrive at
\begin{eqnarray}\label{AB5}
\widetilde{\cal{D}}_{\mu\nu}(x)=
\delta_{\mu\nu}^{\|}\int_{0}^{\infty} ds \
\sqrt{\frac{\pi}{s}}e^{-\frac{x_{4}^{2}}{4s}}\int\frac{d^{3}q}{(2\pi)^{4}}\
e^{-i\mathbf{q}\cdot\mathbf{x}}\ \exp\left(-s\left(
\mathbf{q}^{2}+\frac{2\alpha|eB|}{\pi}e^{-\frac{\mathbf{q}_{\perp}^{2}}{2|eB|}}\right)\right).
\end{eqnarray}
To integrate the $\mathbf{q}$ component, we follow the same steps as
in the previous section [see (\ref{AA6}) and (\ref{AA7})] leading
from (\ref{AA3}) to (\ref{AA8}). In the cylindric coordinates the
photon propagator (\ref{AB5}) is given by
\begin{eqnarray}\label{AB8}
\widetilde{\cal{D}}_{\mu\nu}(x)&=&
\delta_{\mu\nu}^{\|}\int_{0}^{\infty} ds \
\sqrt{\frac{\pi}{s}}e^{-\frac{x_{4}^{2}}{4s}}\int\frac{dq_{\perp}
q_{\perp}dq_{3}d\varphi'}{(2\pi)^{4}}\
e^{-i[q_{\perp}R\sin\theta\cos\left(\varphi-\varphi'\right)+q_{3}R\cos\theta]}\nonumber\\
&&\times\exp\left(-s\left(
\mathbf{q}^{2}+\frac{2\alpha|eB|}{\pi}e^{-\frac{\mathbf{q}_{\perp}^{2}}{2|eB|}}\right)\right).
\end{eqnarray}
Here $\varphi'\in[0,2\pi]$ and
$q_{\perp}\equiv|\mathbf{q}_{\perp}|\in[0,\infty)$. Now using the
integral representation of the Bessel function $J_{0}$ from
(\ref{AA9}) and performing the integration over $\varphi$ we get
\begin{eqnarray}\label{AB10}
\widetilde{\cal{D}}_{\mu\nu}(x)&=&
\delta_{\mu\nu}^{\|}\int_{0}^{\infty} ds \
\sqrt{\frac{\pi}{s}}e^{-\frac{x_{4}^{2}}{4s}}\int\frac{dq_{\perp}
q_{\perp}dq_{3}}{(2\pi)^{3}}\
e^{-iq_{3}R\cos\theta}\ J_{0}\left(q_{\perp}R\sin\theta\right)\nonumber\\
&&\times\exp\left(-s\left(
\mathbf{q}_{\perp}^{2}+q_{3}^{2}+\frac{2\alpha|eB|}{\pi}
e^{-\frac{\mathbf{q}_{\perp}^{2}}{2|eB|}}\right)\right),
\end{eqnarray}
where we have written
$\mathbf{q}^{2}=\mathbf{q}_{\perp}^{2}+q_{3}^{2}$. Performing now
the integration over $q_{3}$ in the same way as the integration over
$q_{4}$ [see (\ref{AB4})] we get first
\begin{eqnarray}\label{AB11}
\int_{-\infty}^{+\infty}dq_{3}\
e^{-sq_{3}^{2}-iq_{3}R\cos\theta}=\sqrt{\frac{\pi}{s}}e^{-\frac{R^{2}\cos^{2}\theta}{4s}},
\end{eqnarray}
and then
\begin{eqnarray}\label{AB12}
\widetilde{\cal{D}}_{\mu\nu}(x)=
\delta_{\mu\nu}^{\|}\int_{0}^{\infty} \frac{ds}{s} \
e^{-\frac{\left(x_{4}^{2}+R^{2}\cos^{2}\theta\right)}{4s}}\int\frac{dq_{\perp}
q_{\perp}}{8\pi^{2}}\ J_{0}\left(q_{\perp}R\sin\theta\right)
e^{-s\left(\mathbf{q}_{\perp}^{2}+\zeta^{2}\exp(-\frac{\mathbf{q}^{2}_{\perp}}{2|eB|})\right)},
\end{eqnarray}
with $\zeta^{2}\equiv\frac{2\alpha|eB|}{\pi}$. To perform the
integration over $\mathbf{q}_{\perp}$, we use the approximation,
$\mathbf{q}_{\perp}^{2}\ll|eB|$, which is valid in the regime of LLL
dominance. After expanding the exponent
$$
e^{-s\left(\mathbf{q}_{\perp}^{2}+\zeta^{2}\exp(-\frac{\mathbf{q}^{2}_{\perp}}{2|eB|})\right)}\simeq
e^{-s\zeta^{2}}e^{-s(1-\frac{\alpha}{\pi})\mathbf{q}_{\perp}^{2}},
$$
the integration over $\mathbf{q}_{\perp}$ can be written as
\begin{eqnarray}\label{AB13}
e^{-s\zeta^{2}}\int_{0}^{\infty} dq_{\perp}\ q_{\perp}\
J_{0}\left(q_{\perp}R\sin\theta\right)
e^{-s\left(1-\frac{\alpha}{\pi}\right)\mathbf{q}_{\perp}^{2}}=
\frac{e^{-s\zeta^{2}}}{2s\left(1-\frac{\alpha}{\pi}\right)}\exp\left(-\frac{R^{2}\sin^{2}\theta}
{4s\left(1-\frac{\alpha}{\pi}\right)}\right).
\end{eqnarray}
To evaluate the $q_{\perp}$ integration, we have used
\cite{gradshteyn}
\begin{eqnarray}\label{AB14}
\int_{0}^{\infty} dz\ z^{\nu+1}J_{\nu}(\beta z)e^{-\gamma
z^{2}}=\frac{\beta^{\nu}}{(2\gamma)^{\nu+1}}e^{-\frac{\beta^{2}}{4\gamma}},\qquad\mbox{Re}(\nu)>-1,\
\mbox{Re}(\alpha)>0,
\end{eqnarray}
by choosing $\gamma=s(1-\frac{\alpha}{\pi})$, $\beta=R\sin\theta$,
and $\nu=0$. Plugging this result in (\ref{AB12}) we arrive at
\begin{eqnarray}\label{AB15}
\widetilde{\cal{D}}_{\mu\nu}(x)=-\frac{1}{16\pi^{2}\left(1-\frac{\alpha}{\pi}\right)}\int_{0}^{\infty}\frac{ds}{s^{2}}\
e^{-s\zeta^{2}-\frac{1}{4s}\left(x_{4}^{2}+R^{2}g^{2}(\theta)\right)},
\end{eqnarray}
where $g(\theta)$ is defined in (\ref{C23}). Defining a new variable
$s'=\zeta^{2}s$, the $s'$-integration can now be performed using
\begin{eqnarray}\label{AB15-a}
\int_{0}^{\infty}\frac{ds'}{s'^{\nu+1}}e^{-s'-\frac{z^{2}}{4s'}}=\left(\frac{2}{z}\right)^{\nu}2K_{\nu}(z).
\end{eqnarray}
We arrive finally at the full LLL photon propagator in the regime
$m_{dyn.}^{2}\ll |\mathbf{q}_{\|}^{2}|\ll|eB|$ is then given by
\begin{eqnarray}\label{AB16}
\widetilde{\cal{D}}_{\mu\nu}(R,\theta,x_{4})=\frac{\delta_{\mu\nu}^{\|}}{4\pi^{2}\left(1-\frac{\alpha}{\pi}\right)}
\frac{\zeta}{\sqrt{x_{4}^{2}+R^{2}g^{2}(\theta)}}K_{1}\left(\zeta\sqrt{x_{4}^{2}+R^{2}g^{2}(\theta)}\right).
\end{eqnarray}
For the notation $x_{4}\to T$ we arrive therefore at our results
from (\ref{C22}).
\end{appendix}


\begin{thebibliography}{99}
\bibitem{new-phases}
T.~Maskawa and H.~Nakajima,
  \textit{Spontaneous symmetry breaking in vector-gluon model},
  Prog.\ Theor.\ Phys.\  {\bf 52}, 1326 (1974); {\it ibid.},
  \textit{Spontaneous breaking of chiral symmetry in a vector-gluon model. 2},
  Prog.\ Theor.\ Phys.\  {\bf 54}, 860 (1975).
\par
  R.~Fukuda and T.~Kugo,
  \textit{Schwinger-Dyson equation for massless vector theory and absence of fermion
  pole},
  Nucl.\ Phys.\  B {\bf 117}, 250 (1976).
\par
  V.~A.~Miransky,
  \textit{Dynamics of spontaneous chiral symmetry breaking and continuum limit in
  Quantum Electrodynamics},
  Nuovo Cim.\  A {\bf 90}, 149 (1985).
\par
  C.~N.~Leung, S.~T.~Love and W.~A.~Bardeen,
  \textit{Spontaneous symmetry breaking in Scale invariant Quantum Electrodynamics},
  Nucl.\ Phys.\  B {\bf 273}, 649 (1986); {\it ibid.},
  \textit{Aspects of dynamical symmetry breaking in gauge field theories},
  Nucl.\ Phys.\  B {\bf 323}, 493 (1989).
\par
  J.~B.~Kogut, E.~Dagotto and A.~Kocic,
  \textit{Catalyzed symmetry breaking in strongly coupled QED},
  Phys.\ Rev.\ Lett.\  {\bf 62}, 1001 (1989).
\bibitem{leung}
C.~N.~Leung, Y.~J.~Ng and A.~W.~Ackley,
  \textit{Schwinger-Dyson equation approach to chiral symmetry breaking in an
  external magnetic field},
  Phys.\ Rev.\  D {\bf 54}, 4181 (1996); {\it ibid.}
  \textit{Chiral Symmetry Breaking by a magnetic field in
  weak-coupling QED}, arXiv:hep-th/9512114.
\bibitem{electron}
  Y.~J.~Ng and Y.~Kikuchi,
  \textit{Narrow $e^+ e^-$ peaks in heavy ion collisions as possibleevidence ofa confinig phase of QED},
  Phys.\ Rev.\  D {\bf 36}, 2880 (1987).
\par
  D.~G.~Caldi and A.~Chodos,
  \textit{Narrow $e^{+}e^{-}$ peaks in heavy ion collisions and a possible new phase of QED},
  Phys.\ Rev.\  D {\bf 36}, 2876 (1987).
\par
  L.~S.~Celenza, V.~K.~Mishra, C.~M.~Shakin and K.~F.~Liu,
  \textit{Exotic States In QED},
  Phys.\ Rev.\ Lett.\  {\bf 57}, 55 (1986).
\par
  D.~G.~Caldi, A.~Chodos, K.~Everding, D.~A.~Owen and S.~Vafaeisefat,
  \textit{Theoretical and phenomenological studies cencerning a possible new phase of QED},
  Phys.\ Rev.\  D {\bf 39}, 1432 (1989).
\par
  D.~G.~Caldi and S.~Vafaeisefat,
  \textit{Chiral symmetry breaking in QED with an external field varying 3-D space
  and time},
  Phys.\ Lett.\  B {\bf 356}, 386 (1995).
\bibitem{miransky1-5}
C.~N.~Leung and S.~Y.~Wang, \textit{Gauge independent approach to
chiral symmetry breaking in a strong magnetic field},  Nucl.\ Phys.\
B {\bf 747} (2006) 266 [arXiv:hep-ph/0510066]; {\it ibid.}, \textit{
Gauge independence and chiral symmetry breaking in a strong magnetic
field,} [arXiv:hep-ph/0503298].
\par
V.~P.~Gusynin, V.~A.~Miransky and I.~A.~Shovkovy, \textit{Large N
dynamics in QED in a magnetic field},
  Phys.\ Rev.\ D {\bf 67}, 107703 (2003)
  [arXiv:hep-ph/0304059].
  \par
V.~P.~Gusynin, V.~A.~Miransky and I.~A.~Shovkovy,
  \textit{Theory of the magnetic catalysis of chiral symmetry breaking in QED},
  Nucl.\ Phys.\ B {\bf 563}, 361 (1999),
  [arXiv:hep-ph/9908320].
\par
  V.~P.~Gusynin, V.~A.~Miransky and I.~A.~Shovkovy,
   \textit{Dynamical chiral symmetry breaking in QED in a magnetic field: Toward
  exact results},
  Phys.\ Rev.\ Lett.\  {\bf 83}, 1291 (1999)
  [arXiv:hep-th/9811079].

\par
  V.~P.~Gusynin, V.~A.~Miransky and I.~A.~Shovkovy,
   \textit{Dimensional reduction and catalysis of dynamical symmetry breaking by a
  magnetic field},
  Nucl.\ Phys.\ B {\bf 462}, 249 (1996)
  [arXiv:hep-ph/9509320].
\par
  V.~P.~Gusynin, V.~A.~Miransky and I.~A.~Shovkovy,
  \textit{Dynamical chiral symmetry breaking by a magnetic field in QED},
  Phys.\ Rev.\ D {\bf 52}, 4747 (1995)
  [arXiv:hep-ph/9501304].
\par
V.~P.~Gusynin, V.~A.~Miransky and I.~A.~Shovkovy,
\textit{Dimensional reduction and dynamical chiral symmetry breaking
by a magnetic field in (3+1)-dimensions}, Phys.\ Lett.\ B {\bf 349},
477 (1995) [arXiv:hep-ph/9412257].
\par
K.~G.~Klimenko,
  \textit{Three-dimensional Gross-Neveu model at nonzero temperature and in an
  external magnetic field},
  Z.\ Phys.\  C {\bf 54}, 323 (1992);   Theor.\ Math.\ Phys.\  {\bf 90}, 1 (1992)
  [Teor.\ Mat.\ Fiz.\  {\bf 90}, 3 (1992)].
\par
S.~Schramm, B.~Muller and A.~J.~Schramm,
  \textit{Quark-anti-quark condensates in strong magnetic fields},
  Mod.\ Phys.\ Lett.\  A {\bf 7}, 973 (1992).
\par
S.~P.~Klevansky and R.~H.~Lemmer,
  \textit{Chiral symmetry restoration in the Nambu-Jona-Lasinio model with constant electromagnetic
  field},
  Phys.\ Rev.\  D {\bf 39}, 3478 (1989).



\bibitem{cond-matter}
  K.~Farakos, G.~Koutsoumbas and N.~E.~Mavromatos,
  \textit{Dynamical flavour symmetry breaking by a magnetic field in lattice
  QED(3)},
  Phys.\ Lett.\  B {\bf 431}, 147 (1998)
  [arXiv:hep-lat/9802037].
\par
  K.~Farakos and N.~E.~Mavromatos,
  \textit{Hidden non-Abelian gauge symmetries in doped planar antiferromagnets},
  Phys.\ Rev.\  B {\bf 57}, 3017 (1998).
\par
  G.~W.~Semenoff, I.~A.~Shovkovy and L.~C.~R.~Wijewardhana,
  \textit{Phase transition induced by a magnetic field},
  Mod.\ Phys.\ Lett.\  A {\bf 13}, 1143 (1998)
  [arXiv:hep-ph/9803371].
\par
  E.~J.~Ferrer, V.~P.~Gusynin and V.~de la Incera,
  \textit{Magnetic field induced gap and kink behavior of thermal conductivity in
  cuprates},
  Mod.\ Phys.\ Lett.\  B {\bf 16}, 107 (2002)
  [arXiv:hep-ph/0101308].
\par
  E.~J.~Ferrer, V.~P.~Gusynin and V.~de la Incera,
  \textit{Thermal conductivity in 3D NJL model under external magnetic field},
  Eur.\ Phys.\ J.\  B {\bf 33}, 397 (2003)
  [arXiv:cond-mat/0203217].
\bibitem{cosmology}
  E.~Elizalde, E.~J.~Ferrer and V.~de la Incera,
  \textit{Neutrino propagation in a strongly magnetized medium},
  Phys.\ Rev.\  D {\bf 70}, 043012 (2004)
  [arXiv:hep-ph/0404234]; {\it ibid.} \textit{Beyond-constant-mass-approximation magnetic catalysis in the gauge
  Higgs-Yukawa model},
  Phys.\ Rev.\  D {\bf 68}, 096004 (2003)
  [arXiv:hep-ph/0209324].
\par
  E.~J.~Ferrer and V.~de la Incera,
  \textit{Neutrino propagation and oscillations in a strong magnetic field},
  Int.\ J.\ Mod.\ Phys.\  A {\bf 19}, 5385 (2004)
  [arXiv:hep-ph/0408108].
\bibitem{shabad-1}
  A.~E.~Shabad and V.~V.~Usov,
  \textit{Modified Coulomb Law in a Strongly Magnetized Vacuum},
  arXiv:0704.2162 [astro-ph]; {\it ibid.}
  \textit{Electric field of a point-like charge in a strong magnetic field},
  arXiv:astro-ph/0607499.
\bibitem{kogut-paper}
  J.~B.~Kogut,
  \textit{A review of the lattice gauge theory approach to Quantum Chromodynamics},
  Rev.\ Mod.\ Phys.\  {\bf 55}, 775 (1983).
\bibitem{schwinger-1}
J.~S.~Schwinger, \textit{On gauge invariance and vacuum
polarization}, Phys.\ Rev.\  {\bf 82}, 664 (1951).
\bibitem{loskutov}
    G.~Calucci and R.~Ragazzon, \textit{Nonlogarithmic terms in the
    strong field dependence of the photon propagator}, J.\ Phys.\ A {\bf
    27}, 2161 (1994).
\bibitem{kuznetsov}
    A.~V.~Kuznetsov and N.~V.~Mikheev,
    \textit{Electron mass operator in a strong magnetic field and
    dynamical chiral symmetry breaking},
    Phys.\ Rev.\ Lett.\  {\bf 89}, 011601 (2002)
  [arXiv:hep-ph/0204201].
\bibitem{ferrer-WI}
  E.~J.~Ferrer and V.~de la Incera,
  \textit{Ward-Takahashi identity with external field in ladder QED},
  Phys.\ Rev.\  D {\bf 58}, 065008 (1998)
  [arXiv:hep-th/9803226].
\bibitem{gradshteyn}
I.~S.~Gradshtein and I.~M.~Ryzhik, {\it Table of integrals, seires
and products}, Academic Press, Orlando (1980).
\bibitem{ng}
  D.~S.~Lee, C.~N.~Leung and Y.~J.~Ng,
  \textit{Chiral symmetry breaking in a uniform external magnetic field},
  Phys.\ Rev.\  D {\bf 55}, 6504 (1997)
  [arXiv:hep-th/9701172].
\bibitem{ferrer}
  E.~J.~Ferrer and V.~de la Incera,
  \textit{Yukawa coupling contribution to magnetic field induced dynamical mass},
  Int.\ J.\ Mod.\ Phys.\  {\bf 14}, 3963 (1999)
  [arXiv:hep-ph/9810473].
\bibitem{rothe}
  H.~J.~Rothe,
  \textit{Lattice gauge theories: An introduction},
  World Sci.\ Lect.\ Notes Phys.\  {\bf 74},  (2005) 1.
\bibitem{wegner}
  F.~J.~Wegner,
  \textit{Duality in generalized Ising models and phase transitions without local
  order parameters},
  J.\ Math.\ Phys.\  {\bf 12} (1971) 2259.
\bibitem{wilson}
  K.~G.~Wilson,
  \textit{Confinement of quarks},
  Phys.\ Rev.\  D {\bf 10}, 2445 (1974).
\bibitem{polyakov}
  A.~M.~Polyakov,
  \textit{Compact gauge fields and the infrared catastrophe},
  Phys.\ Lett.\  B {\bf 59}, 82 (1975).
\bibitem{adler}
  S.~L.~Adler,
  \textit{Photon splitting and photon dispersion in a strong magnetic field},
  Annals Phys.\  {\bf 67}, 599 (1971).
\bibitem{peskin}
  M.~E.~Peskin and D.~V.~Schroeder,
  \textit{An Introduction to Quantum Field Theory}, Reading, USA: Addison-Wesley (1995).

\end{thebibliography}
\end{document}